\documentclass[a4paper,11pt]{article}
\usepackage{jcappub} % for details on the use of the package, please
\usepackage{graphicx}
\usepackage{color}
\usepackage{fontenc}
\usepackage{epsfig}
\usepackage{amsmath, amsthm}
\usepackage[normalem]{ulem}

\newcommand{\be}{\begin{equation}}
\newcommand{\ee}{\end{equation}}
\newcommand{\bea}{\begin{eqnarray}}
\newcommand{\eea}{\end{eqnarray}}

\title{\boldmath  Bulk viscosity coefficients due to phonons in superfluid neutron stars}

\author[a]{Cristina  Manuel,}
\author[b,c]{Jaume Tarr\'us,}
\author[a,d]{Laura Tolos}

\affiliation[a]{Institut de Ci\`encies del Espai (IEEC/CSIC),
Campus Universitat Aut\`onoma de Barcelona,\\
Facultat de Ci\`encies, Torre C5, E-08193 Bellaterra,  Spain
}
\affiliation[b]{Departament d'Estructura i Constituents de la Mat\`eria,
Universitat de Barcelona,\\
Diagonal 647, E-08028 Barcelona, Spain
}
\affiliation[c]{Institut de Ci\`encies del Cosmos,
Universitat de Barcelona,\\
Diagonal 647, E-08028 Barcelona,  Spain
}
\affiliation[d]{Frankfurt Institute for Advanced Studies, 
Johann Wolfgang Goethe University,\\
Ruth--Moufang--Str 1, 60438 Frankfurt am Main, Germany
}
\emailAdd{cmanuel@ieec.uab.es}
\emailAdd{tarrus@ecm.ub.edu}
\emailAdd{tolos@ice.csic.es}

\abstract{
We calculate the three bulk viscosity coefficients  as arising from the collisions among phonons in superfluid neutron stars. We use effective field theory techniques to extract the allowed phonon collisional processes, written as a function of the equation of state of the system. The solution of the dynamical evolution of the phonon number density allows us to calculate the bulk viscosity coefficients as function of the phonon collisional rate and the phonon dispersion law, which depends on the neutron pairing gap. Our method  of computation is rather general, and could be used for different superfluid systems, provided they share the same underlying symmetries. We find that the behavior with temperature of the bulk viscosity coefficients is dominated by the contributions coming from the collinear regime of the $2\leftrightarrow3$ phonon processes.  For typical star radial pulsation frequencies of $\omega \sim 10^{4} s^{-1}$, we obtain that the bulk viscosity coefficients at densities $n \gtrsim 4n_0$ are within $10\%$ from its static value for $T \lesssim 10^9 K$ and for the case of strong neutron superfluidity in the core with a maximum value of the $^3P_2$ gap above 1 MeV, while, otherwise, the static solution is not a valid approximation to the bulk viscosity coefficients. Compared to previous results from Urca and modified Urca reactions, we conclude that at $T\sim 10^9$K phonon collisions give the leading contribution to the bulk viscosities in the core of the neutron stars, except for $n \sim 2n_0$ when the opening of the Urca processes takes place.
}


\begin{document}
\maketitle
\flushbottom

\section{Introduction}

There are strong theoretical reasons to believe that superfluidity occurs in both the inner crust and in the core of neutron stars \cite{Lombardo:2000ec, Heiselberg:1999mq}. The fact that part of the nucleon--nucleon interaction is attractive for some values of the nucleonic density implies the formation of Cooper pairs of neutrons and Cooper pair of protons inside the star, leading to superfluidity and superconductivity, respectively. On the other hand, superfluidity seems to be needed to explain a variety of different neutron star phenomena, such as the existence of pulsar glitches \cite{Anderson:1975zze}, or the cooling of the star \cite{Yakovlev:2004iq}.

Superfluidity would also affect many other aspects of the neutron star dynamics, the corresponding hydrodynamics being essentially different from that of a normal fluid. In particular, rotational and vibrational properties of the star, or the dynamics of the starÕs oscillations, would be different for a star with or without a superfluid core. Further,  the dynamics and damping time scales of both the rotational, vibrational or oscillations modes of the star are governed by the value of the different transport coefficients of the matter inside the star \cite{Cutler90}. In general, one may expect that the value of the different transport coefficients (viscosities, conductivities, etc) might attain much lower values in the superfluid rather than
in the normal phase of the nuclear matter. Several computations of the main transport coefficients in the two phases of nuclear matter can be
found in the literature (see, e.g., Refs.~\cite{Flowers,Cutler87,Haensel:2000vz,Haensel:2001mw,Andersson:2004aa,Shternin:2008es,Alford:2010jf,Manuel:2011ed}).
 
It is  also important to recognize that  the hydrodynamic equations governing the bulk fluctuations of  a superfluid are essentially different from
those of a normal fluid. At non--vanishing temperature one has to employ the two--fluid description of Landau \cite{landaufluids}, which takes into account  the motion of both the superfluid and of the normal component of the system.   In order to describe the different dissipative processes one introduces more transport coefficients than in a normal fluid. In particular, one has three independent bulk viscosities, $\zeta_1, \zeta_2, \zeta_3$, as well as the  shear viscosity and the thermal conductivity \cite{landaufluids,IntroSupe}. The hydrodynamics describing the fluid motions inside a star are even much more complicated, because, apart from the superfluid hydrodynamical equations, one has to describe as well the fluid motions of the charged components (mainly, protons and electrons) of the star.

 The existence of more  bulk viscosity coefficients in superfluid neutron stars has been already noted by Gusakov in Ref.~\cite{Gusakov:2007px}. The computations of Ref.~\cite{Gusakov:2007px} assumed relativistic hydrodynamics instead of the nonÐrelativistic domain that is considered in most of the computations of transport coefficients in the neutron star core, as calculated by Khalatnikov \cite{IntroSupe}, and that we also assume in this manuscript. The first and third bulk viscosity coefficients of Ref.~\cite{Gusakov:2007px} do not have the same dimensions than those of the nonÐrelativistic superfluid hydrodynamics \cite{IntroSupe}. The difference of dimensions of the three bulk viscosity coefficients in Gusakov's approach versus the Landau-Khalatnikov's ones arises due to the fact that they appear in equations of different physical dimensions. The first uses the particle density rather than the mass density, as the hydrodynamical variable. As one might expect that the mass density is the particle density times the mass of the particle (although this is not always the case for strongly interacting systems), there is in principle an easy way to see how the bulk viscosity coefficients introduced by Gusakov are related to those that appear in the Landau-Khalatnikov's hydrodynamical equations (see eqs. (24) and (25) in Ref.~\cite{Gusakov:2007px}). 

%So taken into account this difference of definitions of the bulk viscosity coefficients of Gusakov versus the Landau-Khalatnikov approach, one could obtain the non-relativistic limit of the bulk viscosity coefficients from Gusakov's  equations. We note, however, that this author did not compute the phonon contribution to the bulk viscosity coefficients.}

In this article we compute the superfluid phonon contribution to the three different bulk viscosities in superfluid neutron stars. The phonon contribution to the shear viscosity  in superfluid neutron stars was already computed in Ref.~\cite{Manuel:2011ed,Manuel:2012rd}. The phonon is a collective mode which appears in all superfluids, and which is essential to explain the property of superfluidity, as found out by Landau. In field theoretical language one views the superfluid phonon  as the Goldstone mode which appears due to the fact that the  neutron condensate spontaneously  breaks the $U(1)$ baryonic symmetry.  At low densities, the neutrons pair in a $^1S_0$ channel within the star, while at higher densities a $^3P_2$ channel is preferred  \cite{Lombardo:2000ec}, but the superfluid phonon exists in both of these two different superfluid phases. 
 
The bulk viscosity coefficients depend on a collisional rate of phonon number changing scatterings. Effective field theory (EFT) techniques can be applied to study the phonon self--interactions \cite{Son:2002zn,Son:2005rv}. Then, it is seen that, at leading order (LO) in a derivative expansion, the phonon self--couplings are determined by the equation of state (EoS) of the superfluid. In this paper we consider a simplified model of neutron star made up by neutrons, protons and electrons, using a  causal parametrization of the Akmal,  Pandharipande and  Ravenhall \cite{ak-pan-rav} (APR for short) EoS  \cite{Heiselberg:1999mq} to describe the $\beta$--stable nuclear matter  inside the star, and get from it all the phonon self--couplings. 

Starting from the LO phonon EFT one sees that, in principle, there are several possible phonon number changing processes. However, the energy and momentum conservation laws put kinematical constraints on the possible phonon collisions when corrections to the LO dispersion law
are considered. If the correction to the linear dispersion law curves upward, one phonon can decay into two, but this process is kinematically forbidden in the opposite case. If the correction to the linear dispersion laws curves downward, then the most important process is dominated by $2\leftrightarrow 3$ small angle scatterings, as we will discuss at length. The phonon dispersion law can be computed through a matching procedure with the underlying nucleonic microscopic theory. For neutrons pairing in a $^1S_0$ channel, the phonon dispersion law curves downward  \cite{sarkar}, and in this article we will assume that this is the case in the whole range of densities of the star. Thus we will assume that the collisions determining the values of the
bulk viscosities are $2 \leftrightarrow 3$ phonon scatterings.
 
As the superfluid phonon is a massless mode, one may expect that at sufficiently low $T$ it might give the leading contribution to the  bulk viscosity coefficients. We compare our results for $\zeta_2$ of those obtained from the contribution of both the direct Urca and modified Urca processes \cite{Haensel:2000vz,Haensel:2001mw}, finding that this is indeed the case.

This paper is structured as follows. In Section~\ref{sec:eft} we present the EFT for superfluid phonons at leading order and comment on how corrections at next--to--leading order change the phonon dispersion law. We also present  the nucleonic EoS used in this work, which is a common benchmark for all EoS in neutron star matter. In Section~\ref{sec:bulk} we review the expressions for the phonon contribution to the (static) bulk viscosity coefficients, and get also expressions for the frequency dependent coefficients, while in Section~\ref{sec:rate} we show the phonon decay rate, which consists of $2 \leftrightarrow 3$ collisions, that is relevant for the computation of the bulk viscosity coefficients. Our results for the bulk viscosity coefficients are given in Section \ref{sec:val} and our conclusions in Section~\ref{sec:conc}. Finally, in Appendix~\ref{app:1} we discuss the allowed kinematics when a phonon dispersion law curves downward and in Appendix~\ref{phase-appendix} we give details regarding the phase space integral for the calculation of the phonon decay rate. We use  natural units  $\hbar= c= k_b=1$ in all intermediate computations,  except in the plots as we report our results in CGS units.

\section{ Superfluid phonons interactions and the Equation of State}
\label{sec:eft}

In this Section we first review the Lagrangian describing the  phonon self--interactions using EFT techniques. At LO in a derivative expansion, all the phonon self--couplings can be parametrized in terms of the speed of sound, the density of the superfluid, and the derivatives of the speed of sound with respect to the density at zero temperature $T=0$ (Sec.~\ref{subsec:LO}). This is a result which is valid for any superfluid sharing the same global symmetries. Thus, the phonon physics at LO is determined by the $T=0$ EoS of the superfluid.   In Sec.~\ref{EoSsec} we present the EoS that we use for the superfluid matter in neutron stars for the explicit computations of the bulk viscosity coefficients in the remaining part of this article.  

\subsection{Effective field theory for superfluid phonons at leading order}
\label{subsec:LO}

The superfluid phonon is the Goldstone mode associated to the spontaneous symmetry breaking of a $U(1)$ symmetry, which corresponds to particle number conservation. EFT techniques can be used to write down the effective Lagrangian associated to the superfluid phonon. The effective Lagrangian is then presented as an expansion in derivatives of the Goldstone field, the terms of this expansion being restricted by symmetry considerations. The coefficients of the Lagrangian can be in principle computed from the microscopic theory, through a standard matching procedure, and thus they depend on the short range physics of the system under consideration.

It has been known for a while that the leading--order  term Lagrangian of the Goldstone mode in a superfluid system is entirely fixed by the EoS \cite{Popov}. In recent publications \cite{Son:2002zn,Son:2005rv} it has been realized that at lowest order in a derivative expansion the Lagrangian reads \cite{Son:2005rv}

\begin{eqnarray}
&&\mathcal{L}_{LO}=P\left(X\right)\,,\nonumber \\
&&X=\mu-\partial_t\varphi-\frac{\left(\nabla\varphi\right)^2}{2m}\,,
\end{eqnarray}
where $P(\mu)$ and $\mu$ are the pressure and chemical potential, respectively, of the superfluid at $T=0$. The variable $\varphi$ is the phonon field and $m$ is the mass of the particles that condense. After a Legendre transformation, one can associate this formulation to the one due to Popov \cite{Son:2005rv,Popov}. The associated Hamiltonian has also the same form as the one used by Landau to obtain the self--interactions of the phonons of $^4$He \cite{IntroSupe,Son:2005rv}.

The origin of this particular form for the LO Lagrangian  is that the effective action associated to the theory at its minimum for constant classical field configurations has to be equal to the pressure \cite{Son:2002zn}. This formulation turns out to be very advantageous, as it allows one to derive all the phonon properties at lowest order in a momentum expansion based on the knowledge of the zero temperature pressure of the superfluid. In particular, one can easily get the phonon dispersion law and the form of the leading phonon self--interactions, as well as their leading contribution to different physical processes.

In order to see that, we expand the $P(X)$ around $\mu$, and after the field redefinition
\be
\varphi=\frac{\phi}{\sqrt{\frac{\partial^2P}{\partial \mu^2}}}\,,
\ee
to have the kinetic term canonically normalized, one can write \cite{Escobedo:2010uv}
\begin{eqnarray}
\mathcal{L}_{LO}&&=\frac{1}{2}\left[\left(\partial_t \phi\right)^2-v^2_{ph}\left(\nabla \phi\right)^2\right]-g\left[\left(\partial_t \phi\right)^3-3\eta_g \partial_t \phi \left(\nabla \phi\right)^2\right]  \nonumber \\
&&+\lambda\left[\left(\partial_t \phi\right)^4-\eta_{\lambda,\,1}\left(\partial_t \phi\right)^2\left(\nabla \phi\right)^2+\eta_{\lambda,\,2}\left(\nabla \phi\right)^4\right]+\cdots
\label{lag}
\end{eqnarray}
We have neglected above an irrelevant constant and a total time derivative term, which is only needed to study vortex configurations.

The different  phonon self--couplings of eq.~(\ref{lag}) can be  expressed as different ratios of derivatives of the pressure with respect to the chemical potential \cite{Escobedo:2010uv}. In particular, after using the thermodynamic relation $dP=\frac{\rho}{m}d\mu$, where $\rho$ is the mass density at $T=0$, the phonon velocity is 
\be
v_{ph}=\sqrt{\frac{\frac{\partial P}{\partial \mu}}{m\frac{\partial^2 P}{\partial \mu^2}}}=\sqrt{\frac{\partial P}{\partial \rho}}\equiv c_s \ ,
\ee
that is, it can be identified with  the speed of sound at $T=0$, as it is expected in the low momentum limit. The  dispersion law obtained from this Lagrangian at tree level is exactly  $E_p = c_s p $. 

Defining the quantities
\be
u=\frac{\rho}{c_s}\frac{\partial c_s}{\partial\rho}\,, \quad w=\frac{\rho}{c_s}\frac{\partial^2 c_s}{\partial\rho^2}\,,
\label{precoup}
\ee
we can obtain the three and four phonon self--couplings in terms of the speed of sound, the mass density and derivatives of the speed of sound  with respect to the mass density \cite{Escobedo:2010uv}
\be
\begin{split}
&g=\frac{1-2 u}{6 c_s \sqrt{\rho}}\,,\qquad  \eta_g=\frac{c^2_s}{1-2 u}\,,\qquad \lambda=\frac{1-2 u(4-5u)-2 w \rho}{24c^2_s\rho}\,,\\
&\eta_{\lambda\,,1}=\frac{6c^2_s(1-2 u)}{1-2u(4-5 u)-2w\rho}\,,\qquad  \eta_{\lambda\,,2}=\frac{3c^4_s}{1-2u(4-5 u)-2w\rho} \ .
\label{coup}
\end{split}
\ee
A next--to--leading order (NLO) Lagrangian in a derivative expansion can be constructed as well (see, for example,  the expression of ${\cal L}_{\rm NLO}$ for the cold Fermi gas in the unitarity limit ~\cite{Son:2005rv}). For our purposes, we will not need it, as we will simply compute the leading order $T$ corrections of the bulk viscosity coefficients. It is however relevant for our discussion that at LO the phonon dispersion law is linear in the momentum, and it suffers corrections when one goes beyond the LO expansion. More particularly, at NLO the phonon dispersion law  reads
\be
\label{NLOdisp-law}
%E_p = c_s p + B  p^3   \ ,
E_P = c_s p ( 1 + \gamma p^2) \ .
\ee
The sign of $\gamma$ determines whether the decay of one phonon into two is kinematically allowed or not (see Appendix \ref{kin} for more explicit details). Only dispersion laws that curve upward can allow such processes. The possibility of having these phonon decay processes is important for the value of the different transport coefficients of the superfluid.

The value of $\gamma$ can be computed through a matching procedure with the underlying microscopic theory. For neutrons paring  in a $^1S_0$ channel within neutron stars it can be seen that  \cite{sarkar}
 \be
\gamma=-\ \frac{v_F^2}{45 \Delta^2} , 
\ee
with $v_F$ being the Fermi velocity and $\Delta$ the value of the gap in the $^1S_0$ phase \cite{Lombardo:2000ec}. We will assume that $\gamma$ takes this same value in the $^3P_2$ phase, with $\Delta$ being the angular averaged value of the gap in that phase.  This should be possible to check following the effective field theory techniques of Ref.~\cite{Bedaque:2003wj}. 
Explicit values for the gap function used in this work will be provided in Sec.~\ref{subsec:num}. 

Thus, considering that $\gamma <0$  the first allowed phonon scattering will be binary collisions. In the present work we aim at computing the  bulk viscosities of superfluid neutron star matter. Then,  phonon number changing  processes are needed and the first ones that contribute to the bulk viscosities are $2\leftrightarrow3$ phonon collisions. 

\subsection{Equation of state for superfluid matter in neutron stars }
\label{EoSsec}

The speed of sound at $T=0$ as well as the different phonon self--couplings are determined by the EoS for neutron matter in neutron stars. A common benchmark for a nucleonic equation of state is the one obtained by APR.  Later on  Heiselberg and Hjorth--Jensen \cite{Heiselberg:1999mq} parametrized the APR EoS of nuclear matter in a causal simple form, which will  be subsequently used in this manuscript. The effect of neutron pairing is not considered because it is not expected to have a big impact in the EoS, being typically of order $\Delta^2/\mu^2$, a quantity which remains small.

The parametrized  binding energy per nucleon (E/A) in nuclear matter reads
\begin{eqnarray}
E/A=\mathcal{E}_0 y \frac{y-2-\delta}{1+\delta y}+S_0 y^{\beta}(1-2x_p)^2.
\label{eq1}
\end{eqnarray}
Here $y$ is the ratio of the nucleon particle density ($n$) to nuclear saturation density ($n_0=0.16  \ {\rm fm^{-3}}$),  $y=n/n_0$, and $x_p=n_p/n_0$ is the proton fraction. The nucleon density is given by
\begin{eqnarray}
n=\nu \int^{p_{F}}_0 \frac{d^3p}{(2 \pi)^3} \ ,
\end{eqnarray}
with $p_F$ being the Fermi momentum and $\nu$ the degeneracy factor. In nuclear matter the degeneracy factor $\nu$ is 4. The binding energy per nucleon at saturation density excluding Coulomb energies is $\mathcal{E}_0=15.8 \ {\rm MeV}$ and the parameter $\delta=0.2$ was determined by fitting the energy per nucleon at high density to the EoS of APR \cite{ak-pan-rav} with three-body forces and boost corrections, but taking the corrected values from table 6 of \cite{ak-pan-rav}.  For the symmetry energy at saturation density, Heiselberg et al. obtained $S_0=32 \ {\rm MeV}$ and $\beta=0.6$ for the best fit.

The EoS is given by 
\begin{equation}
\mathcal{E} (n,x_p)=(m+E/A (n,x_p)) \ ,
\label{eq2}
\end{equation}
with $m$ being the mass of the nucleon, while the corresponding nucleonic energy density is 
\begin{eqnarray}
\epsilon_N(n,x_p)=\mathcal{E}(n,x_p) n \ .
\label{eq3}
\end{eqnarray}

For neutron star matter made of neutrons, protons and electrons, the total energy density is the sum of the nucleonic  contribution (neutrons and protons), $\epsilon_N$, and the one for electrons, $\epsilon_e$,
\begin{eqnarray}
\epsilon(n,x_p,n_e)= \epsilon_N(n,x_p)+\epsilon_e(n_e) ,
\end{eqnarray}
with $n_e$ being the density of electrons. The pressure includes also both contributions
\begin{eqnarray}
P(n,x_p,n_e)=P_N(n,x_p)+P_e(n_e) \ ,
\end{eqnarray}
where the nucleonic and electronic contributions to the pressure are given by
\begin{eqnarray}
P_N(n,x_p)&=&\mu_n (n,x_p) \ (1-x_p) \ n + \mu_p (n,x_p) \ x_p n-\epsilon_N(n,x_p), \nonumber \\
P_e(n_e)&=& \mu_e(n_e) n_e-\epsilon_e(n_e)  \ ,
\end{eqnarray}
being $\mu_i$ the chemical potential of each specie. Those chemical potentials are calculated as
\begin{eqnarray}
&&\mu_n(n,x_p)=\frac{\partial \epsilon_N(n,x_p)}{\partial n_n},  \hspace{1cm} 
\mu_p(n,x_p)=\frac{\partial \epsilon_N(n,x_p)}{\partial n_p},  \nonumber \\
&&\mu_e(n_e) \sim p_{F_e} \sim (3 \pi^2 n_e)^{1/3} .
\label{eq:mus}
\end{eqnarray}
where $n_n$ is the density of neutrons, $n_n=(1-x_p)n$.

Nucleons in neutron stars are in  $\beta$--equilibrium against weak decay processes. The constraints imposed  by chemical equilibrium and charge neutrality for matter made of neutrons, protons and electrons are
\begin{eqnarray}
\mu_n&=&\mu_p+\mu_e  \ ,\nonumber \\
\rho_p&=&\rho_e  \ .
\end{eqnarray}
These conditions fix the proportion of protons, neutrons and electrons for each particle density and, thus, the value of the chemical potential, energy density and pressure exerted for each specie at a given density.

For the computation of the speed of sound and the different three and four phonon self--couplings,  we will proceed as follows. The mass density is related to the energy density by the Einstein relation, $\epsilon= \rho$. Then, to compute the mass density we only take into account the nucleonic part,  as $m \gg m_e$. Further, in  $\beta$--equilibrated matter, $n \approx n_n$.  Thus, we will assume that the speed of sound can be computed as
\begin{eqnarray}
c_s(n,x_p)=\sqrt{\frac{\partial P_N(n,x_p)}{\partial \rho_N(n,x_p)}}\approx \sqrt{\frac{\partial P_N(n,x_p)}{\partial n_n} \frac{\partial n_n}{\partial \rho_N(n,x_p)}} , 
\end{eqnarray}
and, for the phonon self--couplings in eq.~(\ref{coup}), we will use the same chain rule.

\section{Phonon contribution to the bulk viscosity coefficients in a generic superfluid }
\label{sec:bulk}

In this Section we review the expressions for the phonon contribution to the (static) bulk viscosity coefficients in a generic superfluid \cite{IntroSupe}. We then generalize these expressions for the situation when there is a periodic perturbation in the system, such that the bulk viscosities depend on the frequency of the perturbation. 

The bulk viscosities  enter as coefficients in  the dissipative hydrodynamic equations.  While the physical meaning of $\zeta_2$ is the same as in a normal fluid, $\zeta_1, \zeta_3, \zeta_4$ refer to dissipative processes which lead to entropy production only in the presence of a space--time dependent relative motion between the superfluid and normal fluid components. The friction forces due to bulk viscosities  can  be  understood as  drops in
the main driving forces acting on  the normal and superfluid components. These forces  are given by the gradients of $P$ and $\mu$, respectively. One can can write that  
\begin{eqnarray}
\label{dissP}
P & =& P_{\rm eq} - \zeta_1 {\rm div}(\rho_s ( {\bf v_n}-{\bf v_s})) - \zeta_2 {\rm div}\,{\bf v_n} \ , \\ \label{dissmu}
\frac{\mu}{m} & = & \frac{\mu_{\rm eq}}{m} - \zeta_3  \, {\rm div}(\rho_s ( {\bf v_n}-{\bf v_s})) - \zeta_4  {\rm div}\,{\bf v_n} \,,
\end{eqnarray}
where $ P_{\rm eq}$ and $ \mu_{\rm eq}$ are the pressure and chemical potential in the absence of bulk viscosities, $ {\bf v_n}$ and ${\bf v_s}$ are the velocities of the normal and superfluid components, respectively, and $\rho_s$ is the mass density of the superfluid component.  There are some fundamental restrictions on the values of these coefficients \cite{IntroSupe}. The Onsager symmetry principle imposes that $\zeta_1= \zeta_4$, and positive entropy production requires that $\zeta_2,\zeta_3 \geq 0$ and that $\zeta_1^2 \leq \zeta_2 \zeta_3$. 

Let us first consider a superfluid system at finite but low temperature $T$, that is slightly away from thermodynamical equilibrium. One can extract the values of the bulk viscosities following a method developed by Khalatnikov \cite{IntroSupe}, which consists of studying the dynamical evolution of the phonon number density $N_{\rm ph}$. It has been shown in Ref.~\cite{Escobedo:2009bh} that this method is equivalent to the method of computing the bulk viscosities using a Boltzmann equation for the phonons in the relaxation time approximation.  

Given $N_{\rm ph}$, we can define $\mu_{\rm ph}$ as the phonon chemical potential. In thermodynamic equilibrium when $\mu_{\rm ph}$ is equal to zero, the number of phonons is a function of the density $\rho$ and the entropy $S$. Let us consider small deviations from equilibrium, for which the density and entropy differ little from their constant equilibrium values. We may, without limiting the generality of the discussion, also consider the velocities 
$ {\bf v_n}$ and ${\bf v_s}$ to be small.  The equation characterizing the approach of the system to equilibrium may be obtained by expanding the rates of change of the phonon numbers, $\partial_t  N_{\rm ph}$, in powers of the chemical potential. If we limit ourselves to terms linear in $\mu_{\rm ph}$, 
%When a perturbation applied to a superfluid system  determines a change of the number of phonons per unit volume, $N_{\rm ph}$, collisional processes tend to restore the equilibrium value of this quantity. 
the evolution equation for the phonon number can be written as 
\be 
\label{Phdenev}
\partial_t  N_{\rm ph} + {\rm div} ( N_{\rm ph} {\bf v_n}) = - \frac{\Gamma_{\rm ph}}{T}   \mu_{\rm ph} \ ,
\ee
 where the rate of change is expressed as a power expansion in the phonon chemical potential  and the decay rate of phonon changing processes, $\Gamma_{\rm ph}$.
 Neglecting quadratic effects, one can still simplify the above equation to
 \be 
\label{Phdenev-linear}
\partial_t  N_{\rm ph} + N_{\rm ph}\, {\rm div}  {\bf v_n} = - \frac{\Gamma_{\rm ph}}{T}   \mu_{\rm ph} \ ,
\ee 
 Expressing the phonon number as a function of the density and of entropy, $N_{\rm ph}(\rho,S)$, and using the linearized continuity hydrodynamic equations for these quantities, one obtains the phonon chemical potential in terms of the different dissipative flows that appear in the hydrodynamic equations  \cite{IntroSupe}. These terms modify  the equilibrium pressure and chemical potential, and with them, and making use of eqs.~(\ref{dissP},\ref{dissmu}), one identifies the different bulk viscosity coefficients.

 Thus, for small departures from equilibrium and for small values of ${\bf v}_s$ and  ${\bf v}_n$ it turns out that~\cite{IntroSupe}
\be
\zeta_i = \frac{T}{\Gamma_{ph}} \, C_i \ , \qquad i=1,2,3,4 \ ,
\label{bulkstatic}
\ee 
where
\be
C_1 = C_4 = -I_1 I_2 \ , \qquad  C_2 = I_2^2 \ , \qquad  C_3 = I_1^2  \ ,
\label{coef}
\ee
and we have defined the quantities $I_1$ and $I_2$ as follows
\bea
I_1 &=&\frac{\partial N_{ph}}{\partial \rho}\,,  \nonumber \\
I_2 &=&N_{ph}-S\frac{\partial N_{ph}}{\partial S}-\rho \frac{\partial N_{ph}}{\partial \rho}\, .
\label{I1I2}
\eea

The quantities $I_1 , I_2$ have been computed in Ref.~\cite{Escobedo:2009bh} for a generic superfluid, realizing that in order to have non--vanishing values of the three coefficients of the bulk viscosities one needs to consider the phonon dispersion law beyond linear order. This  result is in agreement with that one found out by Khalatnikov and Chernikova~\cite{Khalat-Cherni} for the phonons of $^4$He. In Ref.~\cite{Escobedo:2009bh}, and considering the NLO phonon dispersion eq.~(\ref{NLOdisp-law}), it has been found that 
\bea
I_1&=&\frac{60 T^5}{7c^7_s \pi^2}\left(\pi^2\zeta(3)-7\zeta(5)\right)\left(c_s\frac{\partial B}{\partial \rho}-B\frac{\partial c_s}{\partial \rho }\right)\,, \nonumber \\
I_2&=&-\frac{20 T^5}{7c^7_s \pi^2}\left(\pi^2\zeta(3)-7\zeta(5)\right)\left(2Bc_s+3\rho  \left(c_s\frac{\partial B}{\partial \rho}-B\frac{\partial c_s}{\partial \rho }\right)\right)\,,
\label{i1-i2}
\eea
where  $B=c_s \gamma$ and $\zeta(n)$ is the Riemann zeta function.

For astrophysical applications it is however more important to compute the bulk viscosity coefficients when the perturbation that leads the system out  of equilibrium is periodic in time. Then one assumes a time evolution of the phonon density number of the form
\be
N_{ph} =\overline{N}_{\rm ph}+\Re\left(\delta\hat{N}_{\rm ph}e^{i\omega t}\right)\,, 
\label{fluc}
\ee
where $\omega$ is the frequency of the perturbation, $\Re$ denotes the real part, $\overline{N}_{\rm ph}$ stands for the equilibrium value of the phonon density, and $\delta\hat{N}_{\rm ph}$ is the out of equilibrium fluctuation. A similar dependence is assumed for the remaining hydrodynamical variables. It is then easy to generalize the expressions for the transport coefficients in this situation (see for example, Ref.~\cite{Bierkandt:2011zp}). Then the bulk viscosity coefficients turn out to be complex functions which depend on the frequency of the perturbation. Only their real part contributes to the energy dissipation of the system, and this reads
\be
\label{w-bulks}
\zeta_i (\omega) =\frac{1}{1+\left(\omega I^2_1 \,\frac{\partial \rho}{\partial n}\frac{\partial \rho}
{\partial \mu}\frac{T}{\Gamma_{ph}}\right)^2} \frac{T}{\Gamma_{ph}} \, C_i 
  \ , \qquad i=1,2,3,4 \ .
\ee
From the above expressions one can define the value of a characteristic frequency $\omega_c$ for the phonon collisions, defined as 
\be
\omega_c = \frac{1}{I^2_1 \,\frac{\partial \rho}{\partial n}
\frac{\partial \rho} {\partial \mu}}
\frac{\Gamma_{ph}}{T} \  .
\label{char}
\ee
In the limit where $\omega \ll \omega_c$ one recovers the static bulk viscosity coefficients of eqs.~(\ref{bulkstatic}).

In this article we will use the generic expressions showed in this Section to compute the phonon contribution to the (frequency) dependent  bulk viscosity coefficients  for superfluid neutron stars. In such a case, the values  of $\rho, n$ and $\mu$ should correspond to the values of the mass density, particle density and chemical potential of the superfluid neutrons. All the relevant quantities that enter into eqs.~(\ref{w-bulks}) can be  computed with the EoS  and the phonon dispersion law discussed in Sec.~\ref{sec:eft}, except for the phonon decay rate, which we compute in the following Section.

\section{Phonon decay rate}\label{rate}
\label{sec:rate}

In this Section we compute the phonon decay rate relevant  for the computation of the bulk viscosity coefficients. After an expansion or rarefaction of the superfluid, the system  goes back to equilibrium after  a change in the number of particles. Assuming that the phonons are the relevant degrees of freedom in the $T$ regime we consider, this implies that we need to compute a collisional rate of a process that  changes the number of phonons. As discussed at length in Appendix~\ref{kin}, for superfluids where the phonons have a dispersion law with negative values of $\gamma$, the first kinematically allowed scattering consists of $2 \leftrightarrow 3$ collisions. The decay rate associated to these collisions is given by 
\be
\label{rateIntegral}
\Gamma_{ph}=\int d\Phi_5(p_a,\,p_b;\,p_d,\,p_e,\,p_f )\|\mathcal{A}\|^2f(E_a)f(E_b)\left(1+f(E_d)\right)\left(1+f(E_e)\right)\left(1+f(E_f)\right)\,,
\ee
where $f(E)=\left(e^{E/T}-1\right)^{-1}$ is the Bose--Einstein distribution function. 
The phase space is defined as 
\be
\begin{split}
d\Phi_5(p_a,\,p_b;&\,p_d,\,p_e,\,p_f )=\\
&(2\pi)^4\delta^{(3)}\left(\sum_{i=a,b}\vec{p}_i-\sum_{j=d,e,f}\vec{p}_j\right)\delta\left(\sum_{i=a,b}E_i-\sum_{j=d,e,f}E_j\right)\prod_{\substack{k=a,b,\\d,e,f}}\frac{d^3\vec{p}_k}{(2\pi)^3 2E_k}\,,
\end{split}
\ee
and in  Appendix~\ref{phase-appendix} we specify our particular choice of phase space variables. The scattering amplitude $\mathcal{A}$ describes  $2\leftrightarrow 3$ collisions. We compute this rate using the LO  Lagrangian, $\mathcal{L}_{LO}$, given in eq.~(\ref{lag}).  

\begin{figure}
\centering{\includegraphics[width=0.8\textwidth]{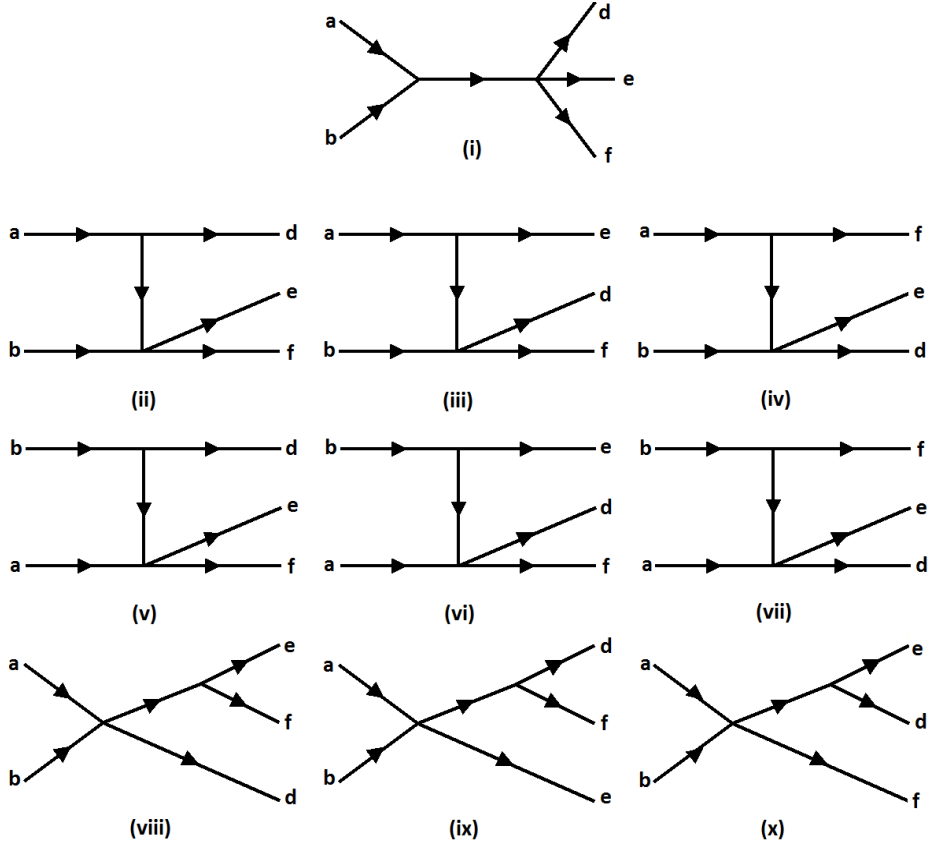}}
\caption{Type I diagrams are formed with one 3--phonon vertex and one 4--phonon vertex of $\mathcal{L}_{LO}.$}
\label{type1}
\end{figure}

\begin{figure}
\centering{\includegraphics[width=0.8\textwidth]{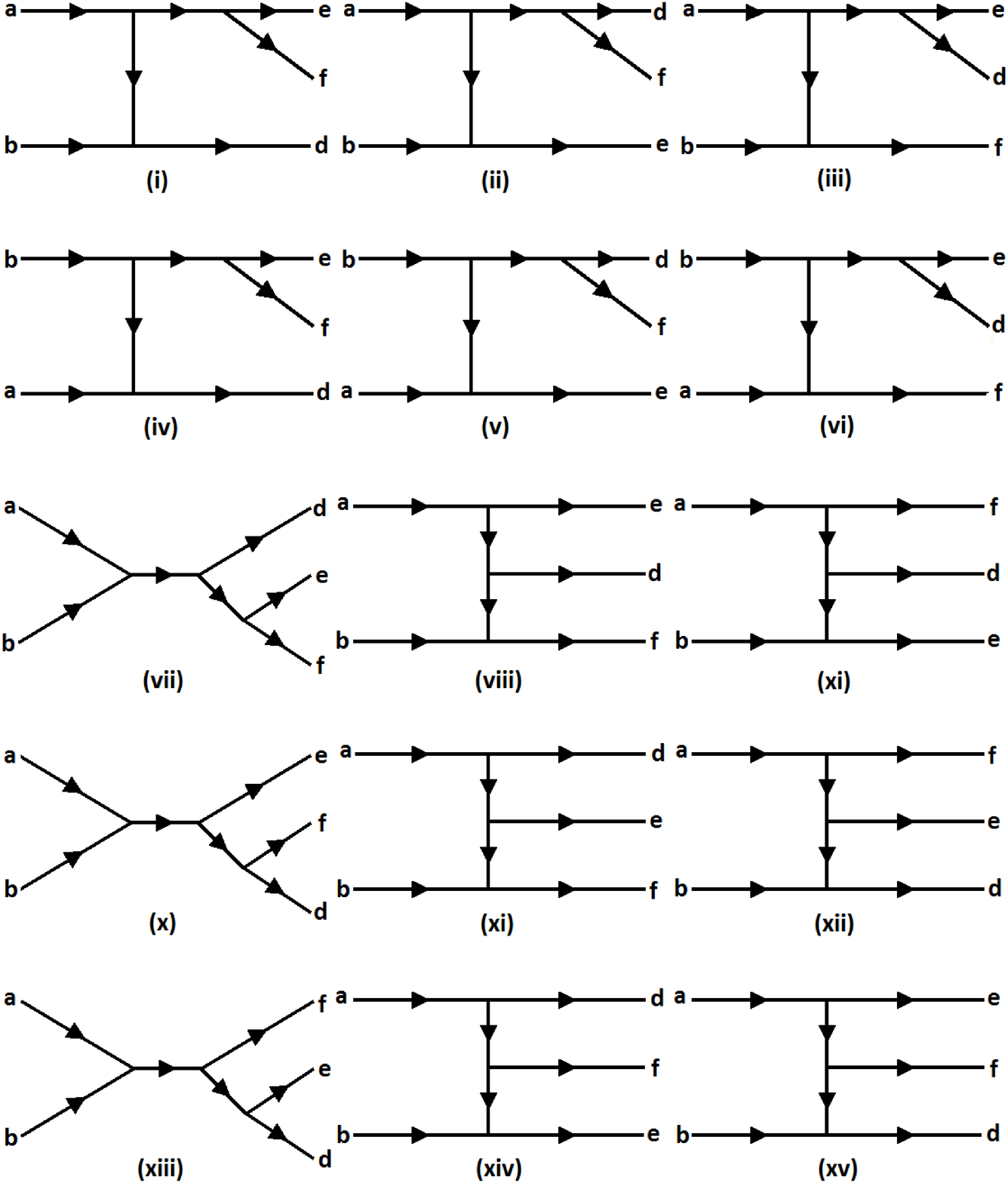}}
\caption{Type II diagrams are constructed with three 3--phonon vertices of $\mathcal{L}_{LO}$ .}
\label{type2}
\end{figure}

The computation is rather involved and has to be done numerically, as one has to consider all the Feynman diagrams depicted in fig.~\ref{type1} and fig.~\ref{type2} to evaluate $\Gamma_{\rm ph}$. We give here some of the details and subtleties associated to the computation, to then present the numerical results obtained for  $\Gamma_{ph}$.

We have classified all possible Feynman diagrams into two groups: those that are formed with one 3--phonon vertex and one 4--phonon vertex of $\mathcal{L}_{LO}$, that we call type I diagrams (see fig.\ref{type1}), and those that are  constructed with three 3--phonon vertices of $\mathcal{L}_{LO}$, (see fig.~\ref{type2}), that we name type II diagrams. The 3--phonon vertex is constructed from the two 3--phonon operators of $\mathcal{L}_{LO}$. Likewise, the 4--phonon vertex is obtained from the contribution of the three 4--phonon operators. The Feynman rules can be found in table~\ref{fr3ph} and table~\ref{fr4ph} for the 3--phonon and 4--phonon operators. The value of $\mathcal{A}$ is obtained after summing the contribution of all the Feynman diagrams depicted in fig.~\ref{type1} and fig.~\ref{type2}, thus $\mathcal{A} = \mathcal{A}^{\text{I}} + \mathcal{A}^{\text{II}}$.

\begin{table}
\centerline{\begin{tabular}{|c||c|c|}\hline
3-phonon & $-g\left(\partial_t \phi\right)^3$ & $3g\eta_g \partial_t \phi \left(\nabla \phi\right)^2$ \\ \hline\hline
\includegraphics[width=2.7cm]{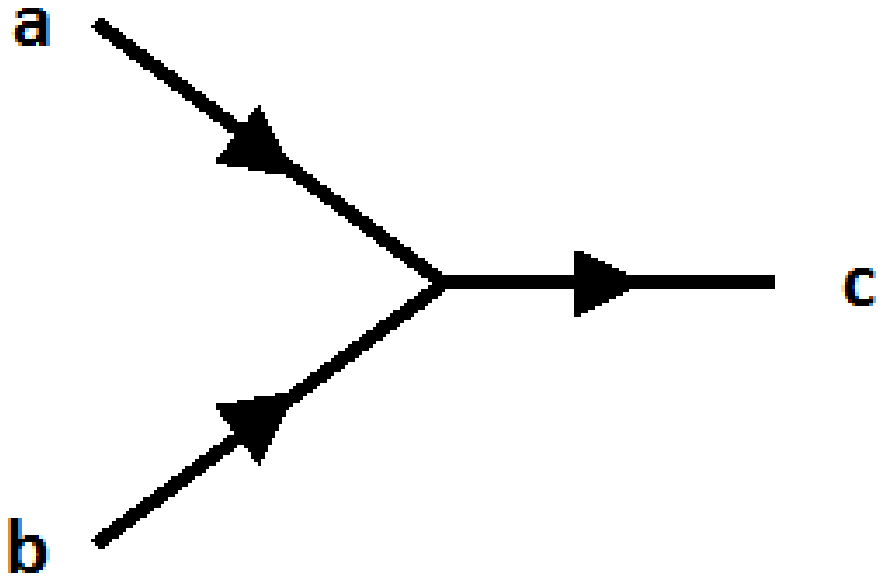} & \raisebox{0.9cm}{$-6gp^0_ap^0_bp^0_c$} & \raisebox{0.9cm}{$18g\eta_gp^0_{\{a}\vec{p}_b\cdot\vec{p}_{c\}}$} \\\hline
\end{tabular}}
\caption{Feynman rules for the 3--phonon operators from the Lagrangian in (\ref{lag}). Note that the specific choice of incoming and outgoing legs depends on the particular diagram. One leg can be changed from incoming to outgoing (and vice--versa) by adding a minus sign. The curly brackets indicate that the quantity has to be symmetrized respect to the indices inside the curly brackets,  by considering the terms coming from all the possible permutations of the indices and dividing by the factorial of the number of indices.}
\label{fr3ph}
\end{table}

\begin{table}
\centerline{\begin{tabular}{|c||c|c|c|}\hline
4-phonon & $\lambda\left(\partial_t \phi\right)^4$ & $-\lambda\eta_{\lambda,\,1}\left(\partial_t \phi\right)^2\left(\nabla \phi\right)^2$ & $\lambda\eta_{\lambda,\,2}\left(\nabla \phi\right)^4$ \\ \hline\hline
\includegraphics[width=2.5cm]{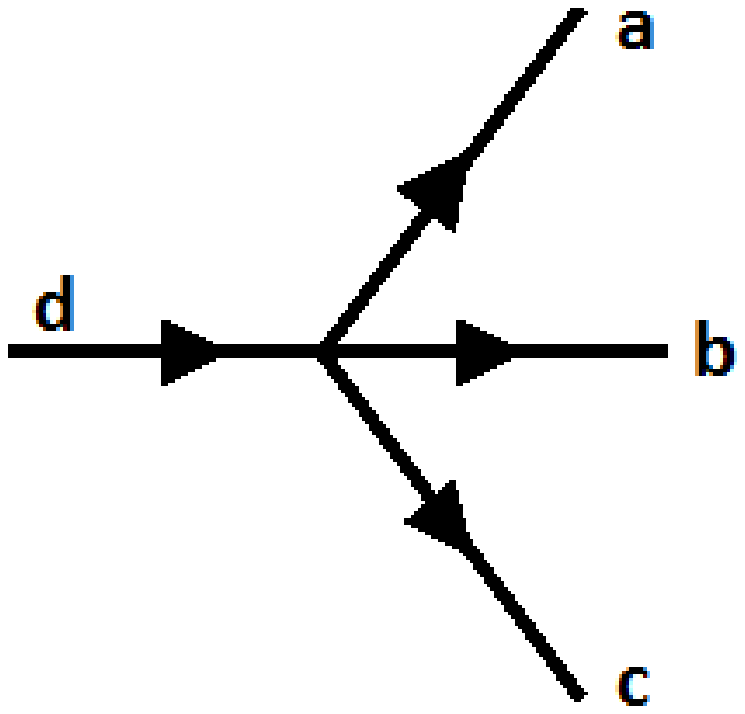} &  \raisebox{0.8cm}{$-i24\lambda p^0_ap^0_bp^0_cp^0_d$} &  \raisebox{0.8cm}{$i24\lambda\eta_{\lambda,\,1}p^0_{\{a}p^0_b\vec{p}_c\cdot\vec{p}_{d\}}$} & \raisebox{0.8cm}{$-i24\lambda\eta_{\lambda,\,1}\vec{p}_{\{a}\cdot\vec{p}_b\vec{p}_c\cdot\vec{p}_{d\}}$} \\ \hline
\end{tabular}}
\caption{Feynman rules for the 4--phonon operators from the Lagrangian in (\ref{lag}). Note that the specific choice of incoming and outgoing legs depends on the particular diagram. One leg can be changed from incoming to outgoing (and vice--versa) by adding a minus sign. The curly brackets indicate that the quantity has to be symmetrized respect to the indices inside the curly brackets, by considering the terms coming from all the possible permutations of the indices  and dividing by the factorial of the number of indices.}
\label{fr4ph}
\end{table}

The scattering amplitude of diagram (i) of fig.~\ref{type1} is given by 
\be
\begin{split}
i\mathcal{A}^{\text{I}}_{\text{(i)}}=&
-\frac{i}{c^3_s\rho^{3/2}}\left[2c^2_s(\vec{p}_a+\vec{p}_b)\cdot p^0_{\{a}\vec{p}_{b\}}-(p^0_a+p^0_b)((1-2u)p^0_a p^0_b-c^2_s\vec{p}_a\cdot \vec{p}_b)\right]\mathcal{G}_{\rm ph}\left(p^0_a+p^0_b,\vec{p}_a+\vec{p}_b\right) \\
&\times\left[\frac{3c^4_s}{2}(\vec{p}_a+\vec{p}_b)\cdot \vec{p}_{\{d} \left(\vec{p}_{e}\cdot\vec{p}_{f\}}\right)-\frac{3c^2_s(1-2u)}{2}\left[(p^0_a+p^0_b)p^0_{\{d}\vec{p}_{e}\cdot \vec{p}_{f\}}+(\vec{p}_a+\vec{p}_b)\cdot p^0_{\{d} p^0_e \vec{p}_{f\}}\right] \right. \\
&\left.+(1+2u(5u-4)-2w\rho)(p^0_a+p^0_b) p^0_d p^0_e p^0_f\right]\,,
\end{split}
\ee
while  for diagram (i) of fig.~\ref{type2} is
\be
\begin{split}
i\mathcal{A}^{\text{II}}_{\text{(i)}}=&\frac{1}{c^3_s\rho^{3/2}}\left[2c^2_s(\vec{p}_b-\vec{p}_d)\cdot p^0_{\{b}\vec{p}_{d\}}-(p^0_b-p^0_d)((1-2u)p^0_b p^0_d-c^2_s\vec{p}_b\cdot \vec{p}_d)\right]\mathcal{G}_{\rm ph}\left(p^0_b-p^0_d,\vec{p}_b-\vec{p}_d\right) \\
&\left[c^2_s(\vec{p}_b-\vec{p}_d)\cdot \left(p^0_a(\vec{p}_e+\vec{p}_f)+(p^0_e+p^0_f)\vec{p}_a\right)-(p^0_b-p^0_d)\left((1-2u)p^0_a (p^0_e+p^0_f)-c^2_s\vec{p}_a\cdot (\vec{p}_e+\vec{p}_f)\right)\right] \\
&\mathcal{G}_{\rm ph}\left(p^0_e+p^0_f,\vec{p}_e+\vec{p}_f\right)\left[2c^2_s(\vec{p}_e+\vec{p}_f)\cdot p^0_{\{e}\vec{p}_{f\}}-(p^0_e+p^0_f)((1-2u)p^0_e p^0_f-c^2_s\vec{p}_e\cdot \vec{p}_f)\right]\,.
\end{split}
\ee
The curly brackets above indicate that the quantity has to be symmetrized with respect to the index inside the bracket. The symmetrization is carried out by considering the terms coming from all the possible permutations of the indices inside the curly brackets and dividing by the factorial of the number of indices inside the curly bracket. $\mathcal{G}_{\rm ph}$ is the phonon propagator
\be
\mathcal{G}_{\rm ph}\left(p^0,\vec{p}\right) = \frac{i}{(p^0)^2 - E_p^2} \ ,
\ee
and at LO, we have $E_p =c_s p$. All the remaining diagrams of fig.~\ref{type1} and fig.~\ref{type2} can be obtained from the expressions given above by relabeling the momenta and using crossing symmetry when necessary. 


Now it is important to realize that for  certain configurations of the momenta the intermediate propagators present in the diagrams on fig.~\ref{type1} and fig.~\ref{type2} will be on--shell, and if one considers that these are computed with a LO dispersion law, this would lead the corresponding amplitude to diverge. This fact has been previously recognized in Ref.~\cite{Mannarelli:2009ia}.

It is possible to determine the configurations of the external momenta that lead the propagator to be on--shell by applying the kinematical considerations associated to the constraints of energy and momentum conservation in every vertex. In all diagrams of both type I and type II the propagators are attached to a 3--phonon vertex with two external legs. When using the linear dispersion law it is easy to prove, see Appendix \ref{kin}, that the propagator is on--shell when the momenta of the 3--phonon vertex are collinear. Therefore, for every diagram, the phase space contains regions where the propagators are on--shell. In type II diagrams, since all verticies are of the 3--phonon type, all external momenta must be collinear. In type I diagrams, depending on the direction of the momentum flow in the propagator, in the 4--phonon vertex we will have either one incoming and three outgoing phonons, or two incoming and two outgoing. When the propagator is on--shell, in the first case all external momenta will be collinear, see Appendix \ref{kin}, but not in the latter one.

The existence of the mentioned collinear singularities in the computation of $\Gamma_{\rm ph}$ is due to the fact that the LO Lagrangian used for its computation is not enough to describe the process under consideration. In order to cure  these divergences, one should also consider the corrections associated to the NLO physics \footnote{It has been claimed that the thermal damping that appears at one-loop might also cure the collinear singularity \cite{Manuel:2004iv}. However, at NLO, the imaginary part of the phonon self-energy vanishes when evaluated on--shell for $\gamma < 0$ \cite{Escobedo:2010uv}, and thus cannot regulate the collinear singularity in this case.}. We will not take into account NLO corrections to the different 3 and 4--phonon vertices, as following the power counting associated to the phonon EFT, it is easy to realize that these would only represent $T^2/\Delta^2$ corrections to our results. When the phonon  propagator is considered with a NLO dispersion law, and taking into account that $\gamma <0$, then energy and momentum conservation applied to every vertex of the different Feynman diagrams allows to deduce that there is no configuration of the external momenta that makes the internal phonon propagator to be on--shell (for details see Appendix~\ref{kin}). 

In the computation of $\Gamma_{\rm ph}$ we will thus consider all the Feynman diagrams of fig.~\ref{type1} and fig.~\ref{type2} using the phonon propagators with a NLO dispersion law. One then sees that, in this case, the almost collinear region (or small angle scattering region) of the available phase space is enhanced with respect to the rest of the phase space (or large angle scattering region) because the denominator of the phonon propagators in this region is small. In particular, this can be seen if we write the NLO phonon propagator as
\be
\mathcal{G}_{\rm ph}\left(p^0_i+p^0_j,\vec{p}_i+\vec{p}_j\right)=i \left[c^2_s(p_i+p_j)^2\left[1+2\gamma\left(\frac{p_i^3+p_j^3}{p_i+p_j}\right)\right]-c^2_s \left(\vec{p}_i+\vec{p}_j\right)^2\left[1+2\gamma \left(\vec{p}_i+\vec{p}_j\right)^2\right]\right]^{-1}\,.
\label{prop}
\ee
Then, it is possible to see that in the almost collinear region, where  the angle between $\vec{p}_i$ and $\vec{p}_j$ is $\theta_{ij}\approx0$, the propagator behaves as $\sim 1/p^4$, as compared to the region of large angle scattering, where the propagator behaves as $\sim 1/p^2$. As a result, one can easily deduce that the almost collinear region of the phase space is enhanced with respect to the rest of the phase space. 

Since the NLO terms of the propagator are only relevant in the almost collinear region, which is defined by $\cos(\theta_{ij})\sim 1$, 
we will approximate these terms by using the first order contribution in an expansion about $\cos(\theta_{ij}) = 1$. This leads to the following, much simpler approximated expression of the propagator of eq.~(\ref{prop}), which we have used in our explicit numerical calculations:
\be
\mathcal{G}_{\rm ph}\left(p^0_i+p^0_j,\vec{p}_i+\vec{p}_j\right) \approx i \left[2c^2_s p_i p_j\left(1-\cos(\theta_{ij})-3\gamma(p_i+p_j)^2 \right)\right]^{-1}\,.
\label{propx}
\ee
Let us finally comment that some simple dimensional analysis allows us to see how the decay rate $\Gamma_{\rm ph}$ associated to large or small angle scatterings behave \cite{Mannarelli:2009ia}. We can obtain the dependence on $T$ of the rate of the phonon number changing process by defining dimensionless variables associated to the momenta, $x_i=\frac{c_s p_i}{T}$. Using these dimensionless variables, all the $T$ dependence of the integral in eq.~(\ref{rateIntegral}) factorizes. For the phase space $d\Phi_5$ the dependence is $T^6$ (see Appendix~\ref{phase-appendix}). The $T$ dependence of the amplitude corresponding to the diagrams in fig.~\ref{type1} and fig.~\ref{type2} can be computed by summing the number of legs of each one of the vertices of the diagram, which amounts to the total number of derivatives and hence the powers of the external momenta. From this number we have to subtract the momentum dependence of the phonon propagator(s). The $T$ dependence of the rate is then, two times the dependence of the amplitude added to the dependence of the phase space. For large angle collisions, then one sees that 
\be
\Gamma_{\rm ph} \propto T^{16}\,.
\ee
In the almost collinear region the dependence of the propagators on the momentum is $p^4$ instead of $p^2$. Thus, the contributions to the rate coming from the collinear region have a different dependence on $T$. The $T$ dependence of the collinear regions depends if we compute $\Gamma_{\rm ph}$ with only the scattering matrix of the type I diagrams, or only with the scattering matrix of type II diagrams, or with the cross terms (that is, using $\|\mathcal{A}^{\text{I}} \mathcal{A}^{\text{II}}\|$ in $\Gamma_{\rm ph}$):
\be
\Gamma^{\text{type I}}_{\rm ph}  \propto  T^{12} \,, \quad \Gamma^{\text{type II}}_{\rm ph} \propto  T^{8} \,, \quad \Gamma^{\text{cross}}_{\rm ph} \propto  T^{10} \,. 
\ee 

\subsection{Numerical results for the phonon decay rate}
\label{subsec:num}

\begin{figure}
\centering{\includegraphics[width=8cm]{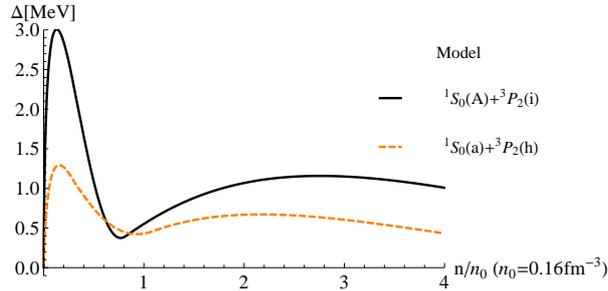}}
\caption{The sum of the $^1S_0$ and angle-averaged $^3P_2$ neutron gaps as a function of the nucleon particle density in units of saturation density, $n_0=0.16 \ {\rm fm^{-3}}$.  Two different models for the sum of the $^1S_0$ and angle-averaged $^3P_2$ neutron pairing gaps  have been considered: a) $^1S_0 (A)$$+$$^3P_2 (i)$ model, where  the $^1S_0$ neutron gap  is calculated in the BCS approach using different bare nucleon--nucleon interactions that converge towards a maximum neutron gap of about 3 MeV at $p_F \approx 0.85 {\rm fm}^{-1}$ (parametrization $A$ of table I in Ref.~\cite{Andersson:2004aa}) while for $^3P_2$ we have taken the parametrization $i$ (strong neutron superfluidity in the core); b) $^1S_0 (a)$$+$$^3P_2 (h)$ model, where the $^1S_0$ neutron gap  incorporates medium polarization effects (parametrization $a$), whereas for the $^3P_2$ neutron gap we have considered the parametrization $h$ (strong neutron superfluidity).}
\label{gap}
\end{figure}

\begin{figure}
\centerline{\includegraphics[width=8cm,height=6cm]{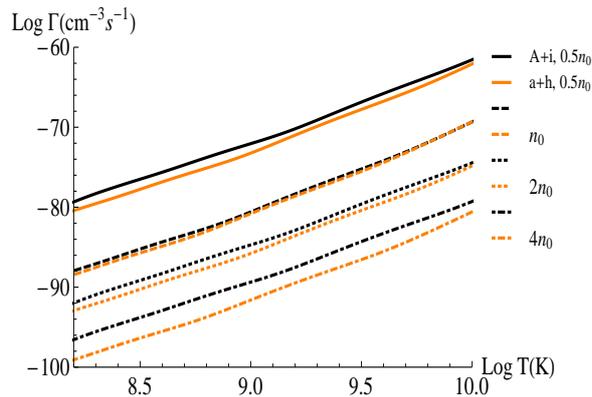}}
\caption{Phonon rate in CGS units as a function of the temperature for different densities and the two neutron gap models.}
\label{rates}
\end{figure}

The evaluation of the phonon rate depends on the features of the EoS as well as the value and density dependence of the neutron pairing gap in the core of neutron stars. While the APR nucleonic EoS is a common benchmark for all the EoS used in neutron star matter, it is still under debate the exact values of the  $^1S_0$ and $^3P_2$ neutron gaps and their density dependence \cite{Lombardo:2000ec}. It is, though, believed that $^1S_0$ neutron gap extends up to $n \approx n_0/2$ while neutrons are gapped in $^3P_2$ well inside the core of the neutron star. 

%{\blue The sum of the $^1S_0$ and  angle-averaged $^3P_2$ neutron gaps used throughout this work are shown in fig.~\ref{gap}. The use of the angle-averaged $^3P_2$ neutron gap is a common practice that, however, can underestimate the contribution of the phonon interactions to the bulk viscosity due to the fact that ignores the spontaneous symmetry breaking of the rotational symmetry, and in this case there are additional
%massless collective excitations. The Goldstone modes associated to the breaking of rotational invariance, the so-called
%angulons \cite{Bedaque:2003wj}, should as well contribute to the flavor changing processes and to the bulk viscosity. Thus, our results using the angle-averaged $^3P_2$
%neutron pairing should be considered as a lower bound.}

In fig.~\ref{gap} we have considered  two very different gap models as a function of the density in order to illustrate the model dependence of our results. Our first model, named hereafter $^1S_0(A)$$+$$^3P_2(i)$,  consists of the $^1S_0$ neutron gap that results from the BCS approach using different bare nucleon--nucleon interactions that converge towards a maximum gap of about 3 MeV at $p_F \approx 0.85 {\rm fm}^{-1}$ (parametrization $A$ of table I in Ref.~\cite{Andersson:2004aa}). The anisotropic $^3P_2$ neutron gap is more challenging and not fully understood as one must extend BCS theory and calculate several coupled equations while including relativistic effects since the gap extends for densities inside the core. We have taken the parametrization $i$ (strong neutron superfluidity in the core) of table I in Ref.~\cite{Andersson:2004aa} for the $^3P_2$ neutron angular averaged value, which presents a maximum value for the gap of approximately 1 MeV. The second model considered, $^1S_0(a)$$+$$^3P_2(h)$, goes beyond BCS for the $^1S_0$ neutron gap as it incorporates medium polarization effects (parametrization $a$). The maximum value for the gap is then reduced to 1 MeV.  Moreover, for the $^3P_2$ neutron gap we have taken into account the parametrization $h$ (strong neutron superfluidity) with a maximum value of about 0.5 MeV. We have, though, not considered weak neutron superfluidity in the core, as discussed in Ref.~\cite{Andersson:2004aa}. In this weak superfluid regime the values of the gap are $\Delta\lesssim$ 0.1 MeV for densities well inside the core. Thus, the corresponding transition temperatures from the superfluid to the normal phase are $T_c \sim 1/2 \Delta \lesssim 5 \times 10^8 K$ and no superfluidity is expected in the neutron star core for the temperatures studied.

The phonon decay rate as a function of the temperature is displayed in fig.~\ref{rates} for densities from $0.5n_0$ to $4n_0$ and for the two gap models previously discussed. These rates have been obtained  solving  eq.~(\ref{rateIntegral}) numerically by means of Montecarlo integration. We can gauge the overall dependence of the rates with temperature by fitting a function of the form $C\, T^{m}$, where $C$ and $m$ are the parameters to be determined. The rates scale with $T \sim 10^{10-11} $K, the exact value depending on the density and the neutron gap model used. Given the previous dimensional analysis for the temperature dependence of the type I, type II and cross contributions to the rate, we conclude that the collinear region dominates over the non--collinear regime.

\begin{figure}
\includegraphics[width=8cm,height=6cm]{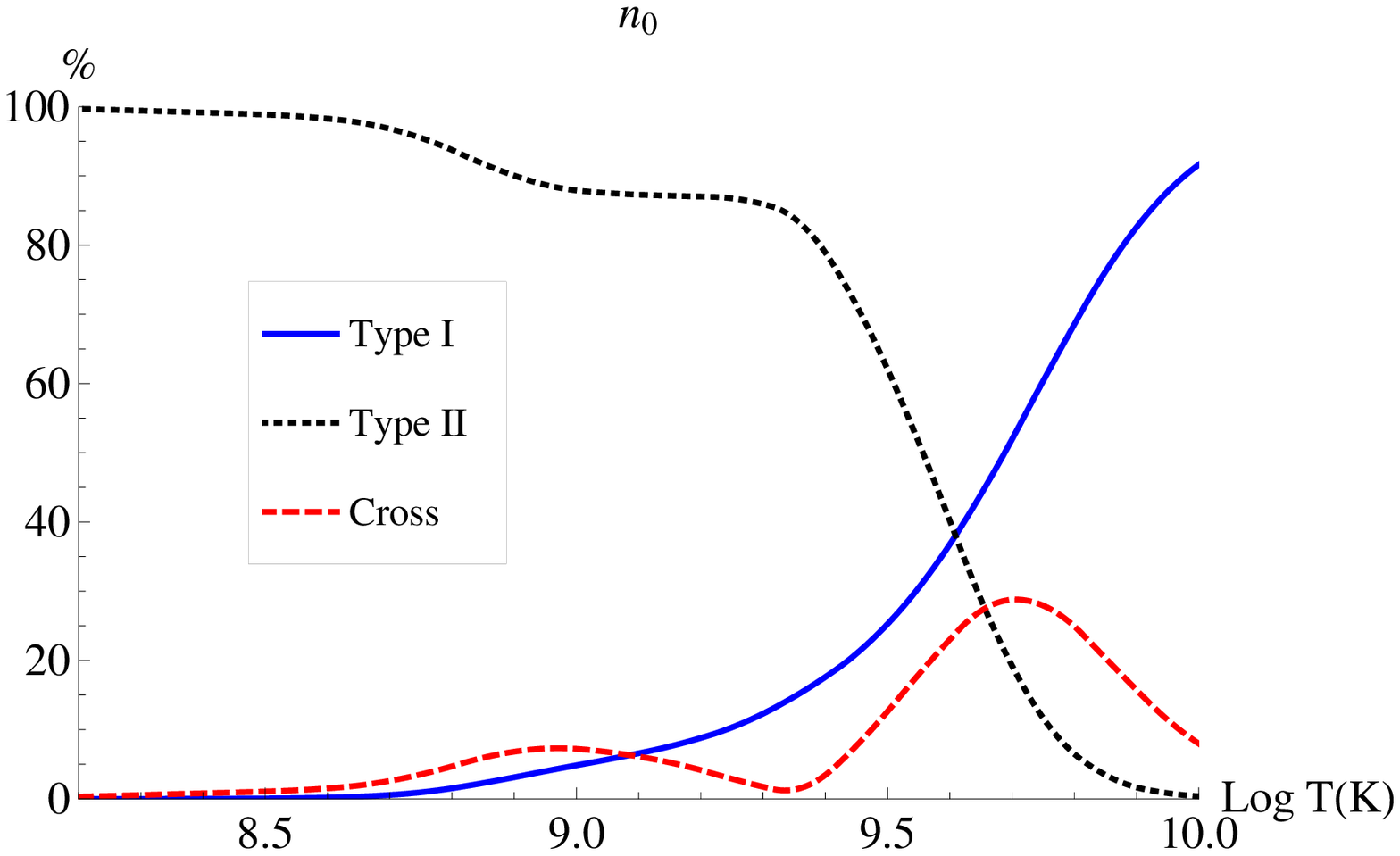}
\hfill
\includegraphics[width=8cm,height=6cm]{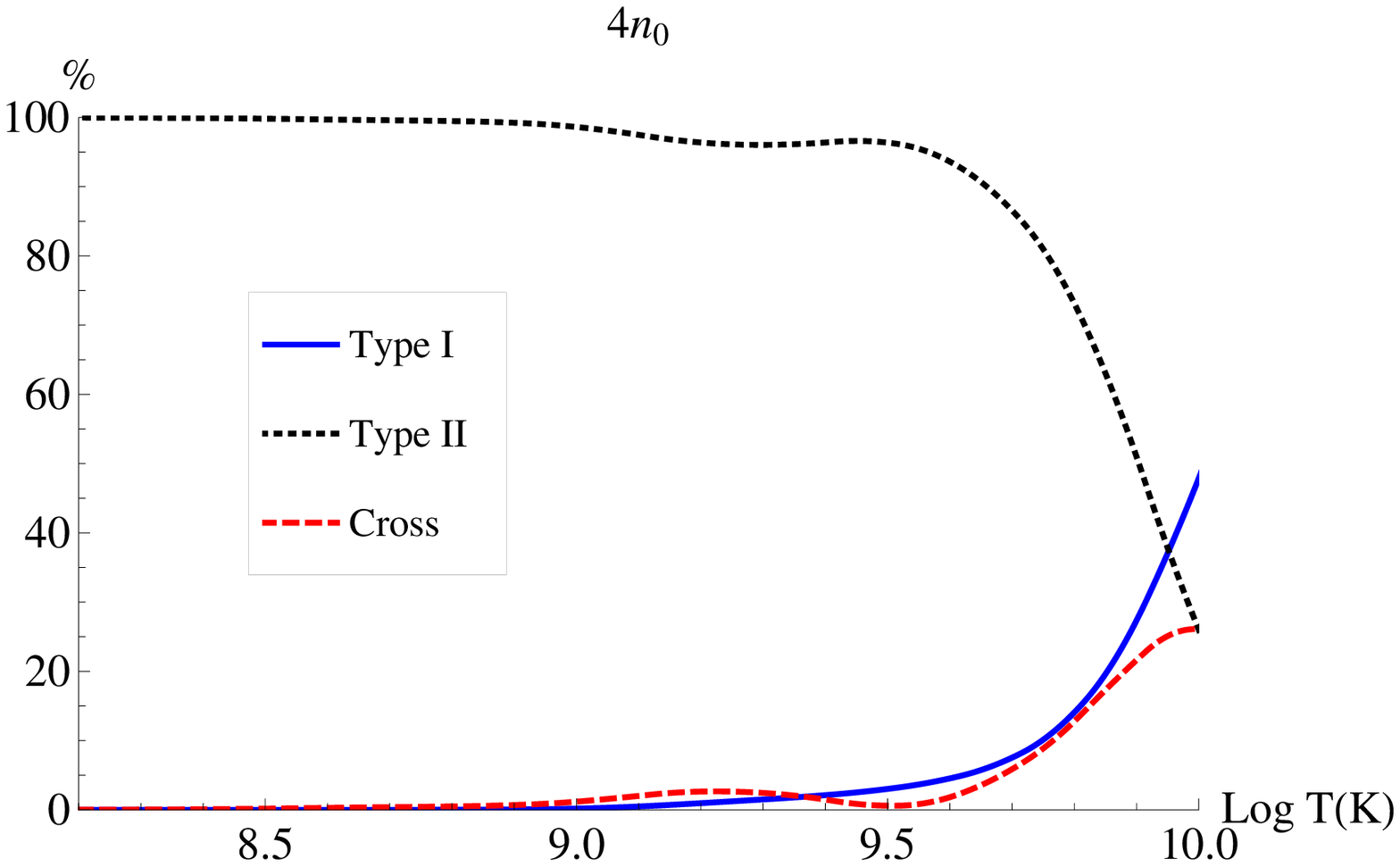}
\caption{The relative contribution of type I, type II and cross terms to the phonon rate as a function of the temperature for $n=n_0$ (left panel) and $n=4n_0$ (right panel). We use model $^1S_0 (A)$$+$$^3P_2 (i)$ for the neutron pairing gap. A similar behavior of these contributions is obtained for the $^1S_0 (a)$$+$$^3P_2 (h)$ neutron gap model.}
\label{pccon}
\end{figure}

Moreover, the relative importance of the type I, type II and cross contributions is shown in fig. \ref{pccon} for $n=n_0$ and $n=4n_0$ using the  $^1S_0(A)+^3P_2(i)$ neutron gap model. We observe that type II diagrams govern the behavior of the phonon decay rate up to $T\sim 10^{9}$K from $n_0$ to $4n_0$. As density increases, the dominance of type II terms extends to higher temperatures. Type I terms are sizeable as the temperature approaches to $T \sim 10^{9.5-10}$K, specially for $n=n_0$, whereas the cross contributions remain small but non--negligible as temperature augments. A similar behaviour is obtained for the type I, type II and cross contributions to the rate using the $^1S_0(a)+^3P_2(h)$ neutron gap model.

\section{Values of the bulk viscosity coefficients} 
\label{sec:val}

The static and frequency--dependent bulk viscosity coefficients of eqs.~(\ref{bulkstatic}) and eqs.~(\ref{w-bulks}), respectively, result, on one hand, from the computation of the phonon rate and, on other hand, from the calculation of the $I_1$ and $I_2$ terms. The quantities $I_1 , I_2$  depend on the EoS and, in particular, on the value and density dependence of the neutron pairing gaps (eqs.~\ref{i1-i2}).  Thus, the exact numerical results for the bulk viscosity coefficients calculated in the following will unavoidably depend on the chosen value of the neutron gaps and its density dependence, although the method of computation itself is rather general. 

\begin{figure}
%\begin{tabular}{ccc}
\includegraphics[width=8cm,height=6cm]{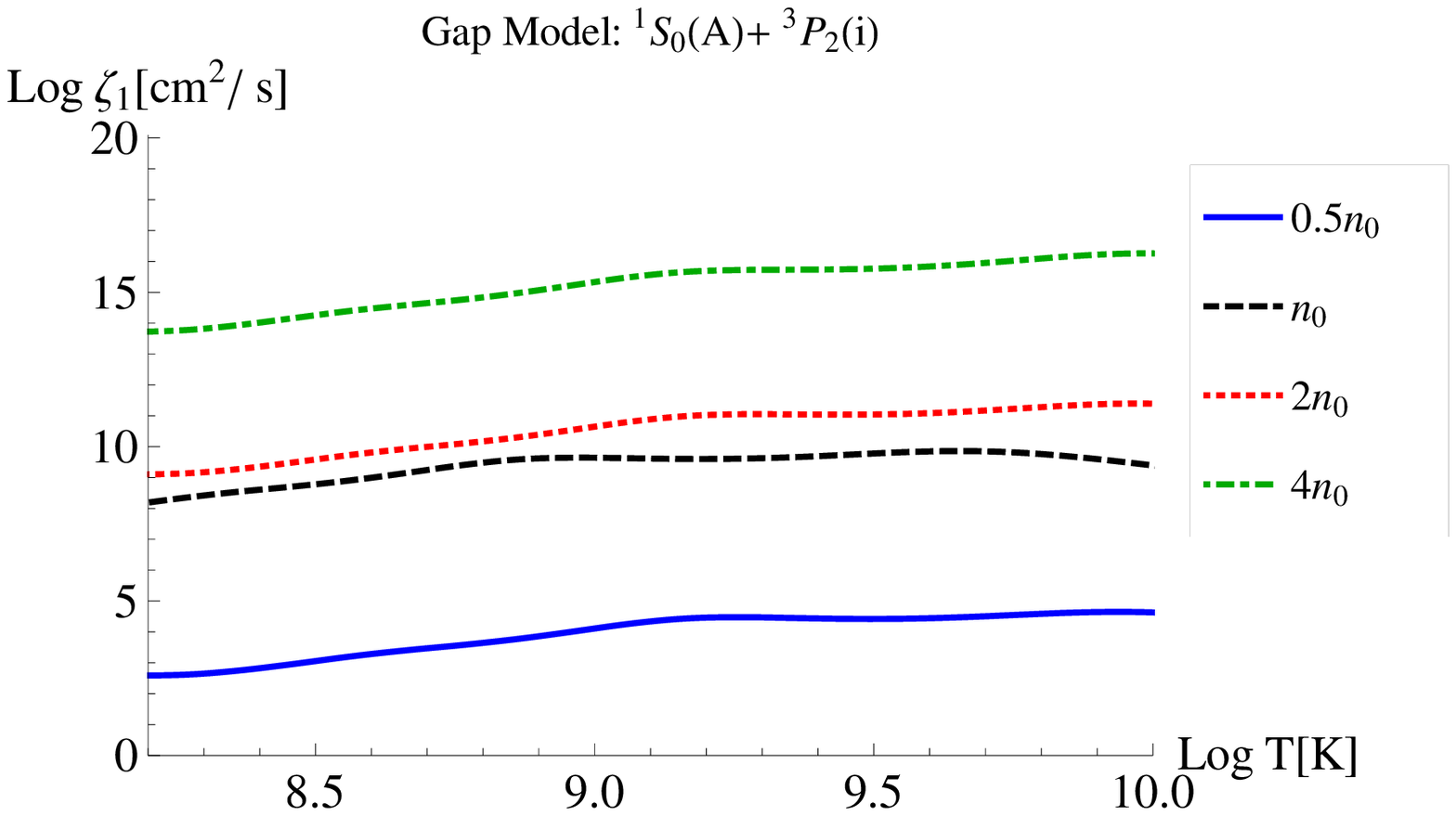} 
\hfill
\includegraphics[width=8cm,height=6cm]{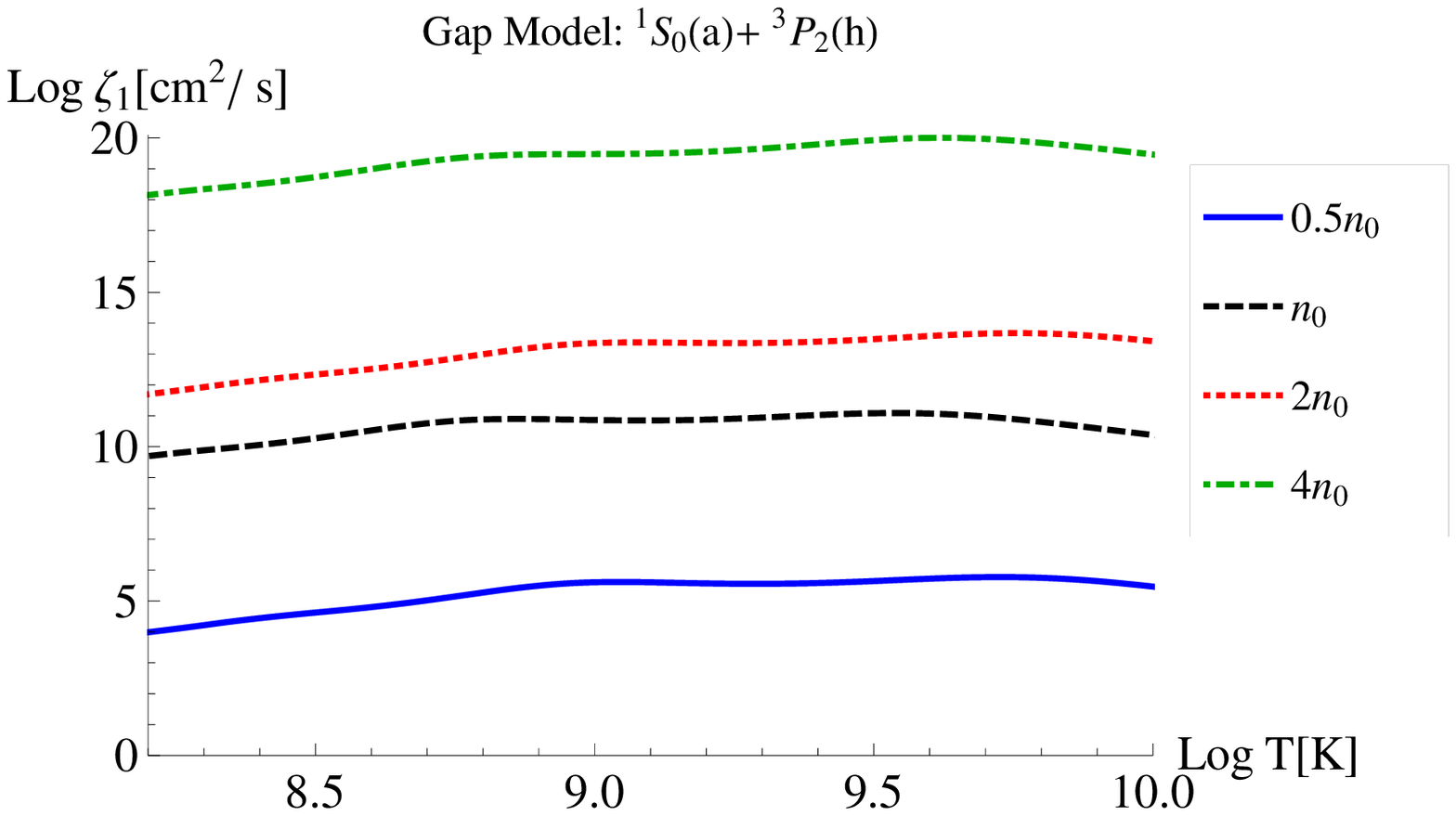}  
%\end{tabular}
\caption{$\zeta_1$ static bulk viscosity coefficient  as a function of the temperature for various densities for the two neutron gap models.}
\label{stvis1}
\end{figure}

\begin{figure}
%\begin{tabular}{ccc}
\includegraphics[width=8cm,height=6cm]{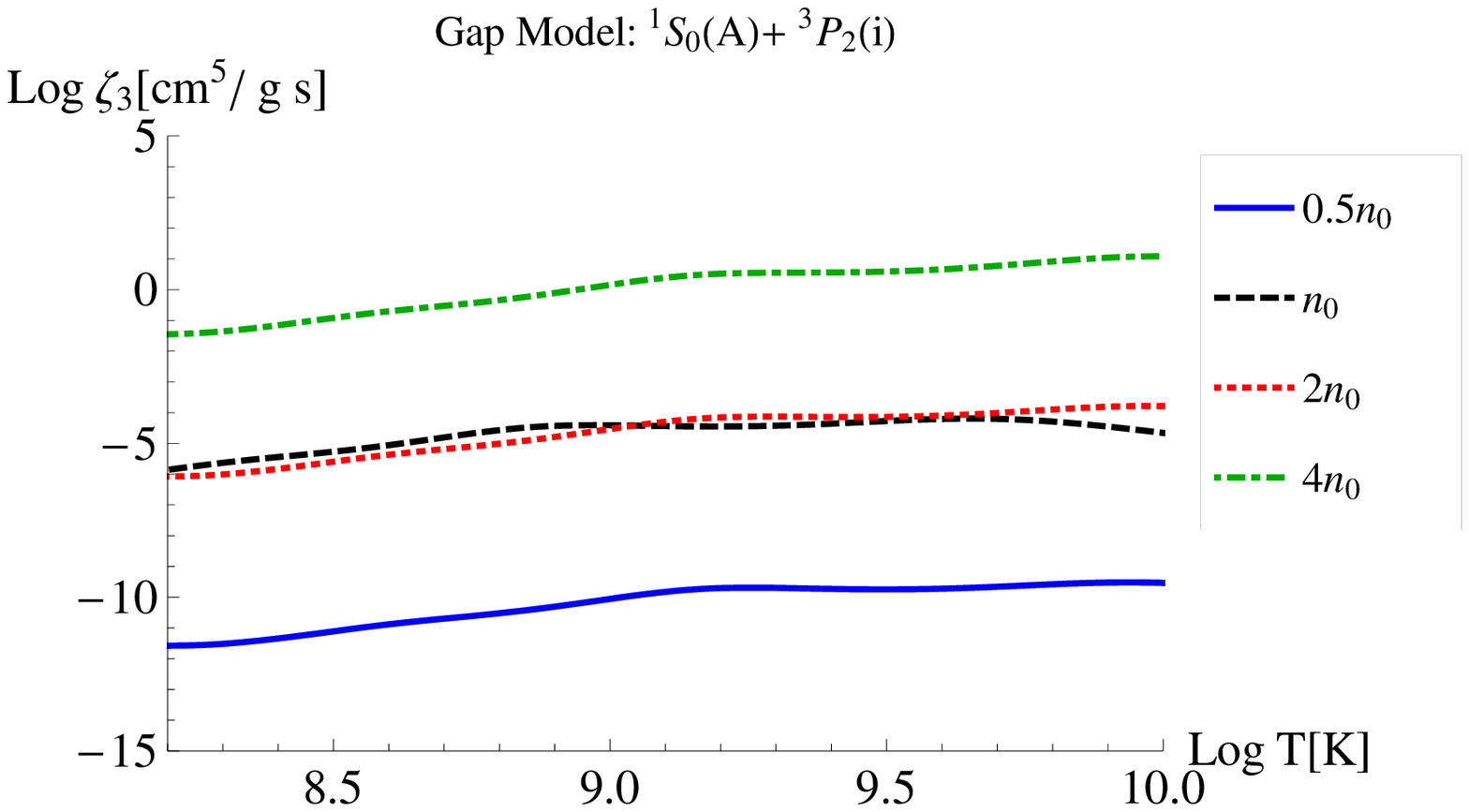} 
\hfill
\includegraphics[width=8cm,height=6cm]{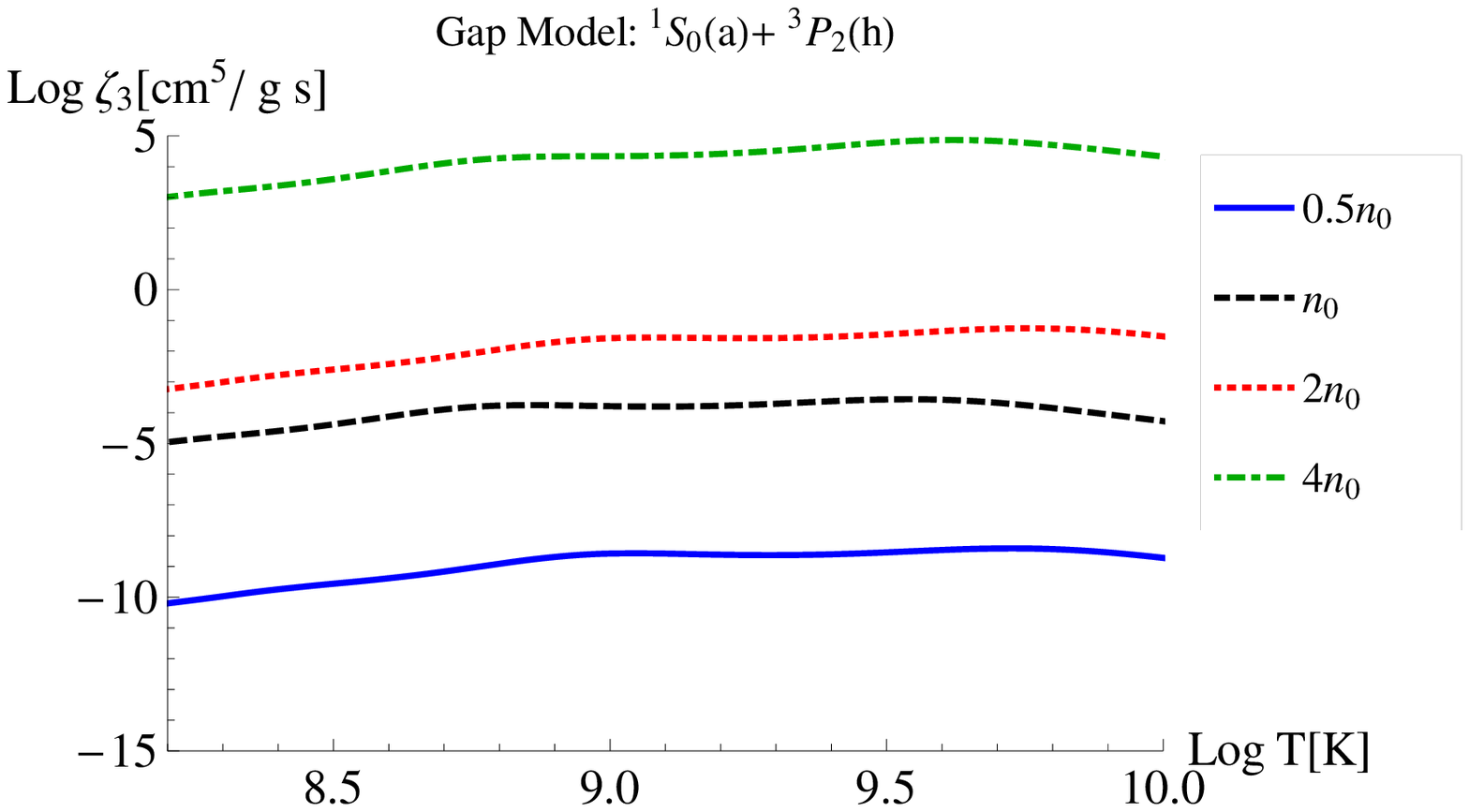}  
%\end{tabular}
\caption{$\zeta_3$ static bulk viscosity coefficient  as a function of the temperature for various densities for the two neutron gap models.}
\label{stvis3}
\end{figure}

\begin{figure}
\begin{tabular}{ccc}
\includegraphics[width=8cm,height=6cm]{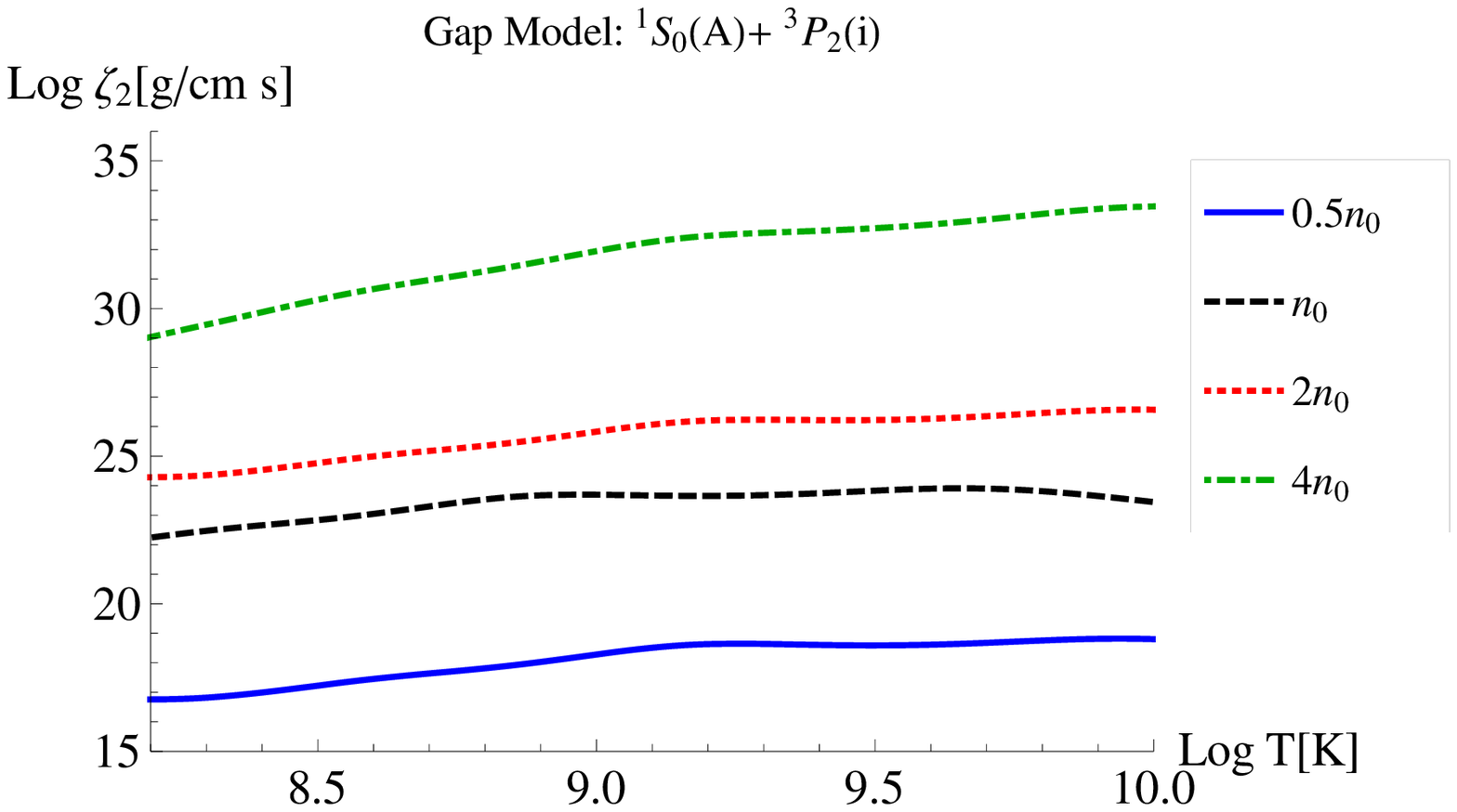} & \includegraphics[width=8cm,height=6cm]{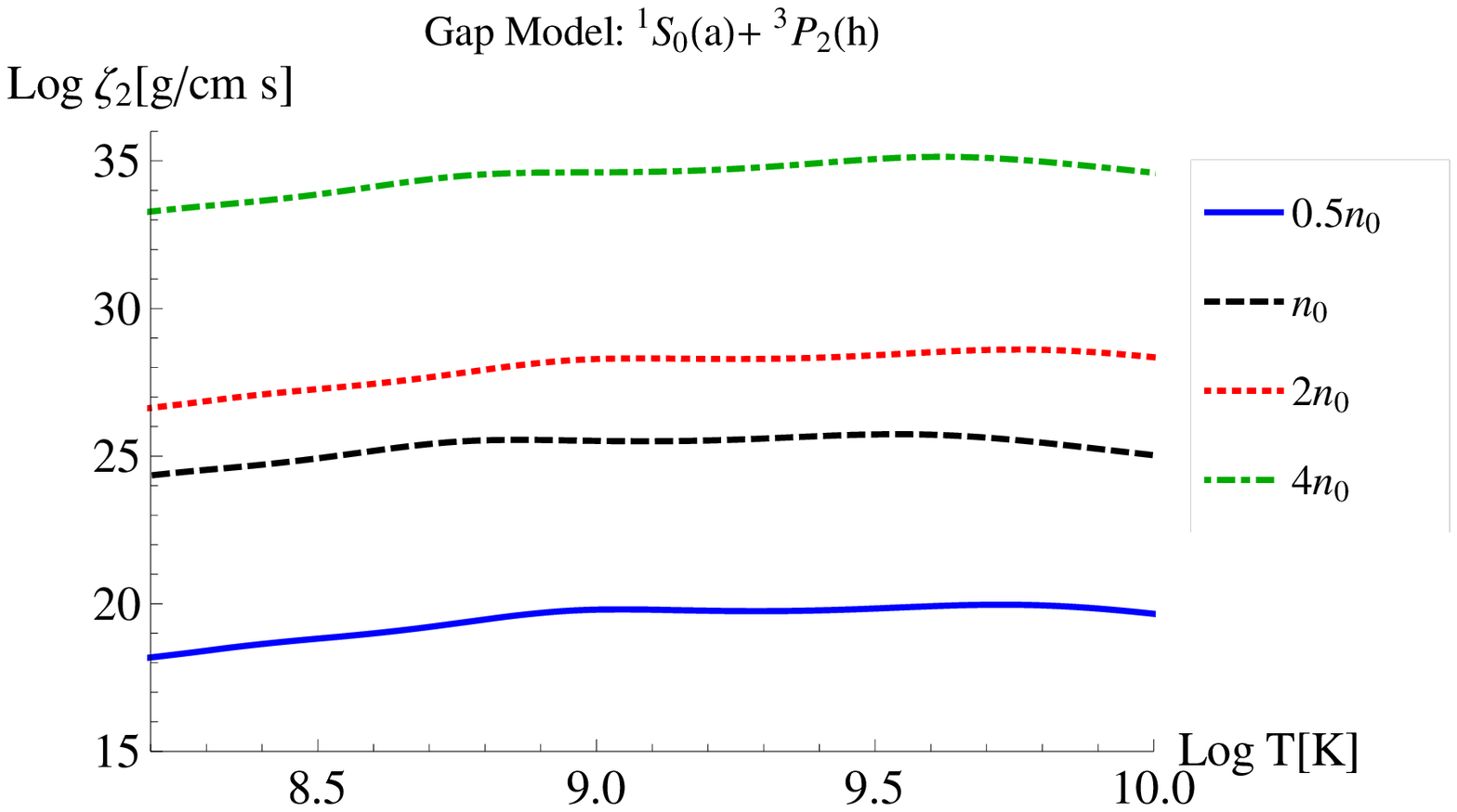}
\end{tabular}
\caption{ $\zeta_2$ static bulk viscosity coefficient as a function of the temperature for various densities for the two neutron gap models.}
\label{stvis2}
\end{figure}

In figs.~\ref{stvis1},\ref{stvis3} and \ref{stvis2} we present the static $\zeta_1$, $\zeta_3$ and $\zeta_2$ bulk viscosity coefficients as a function of the temperature for densities between $0.5n_0$ and $4n_0$.  The temperature range corresponds to the typical temperatures used in neutron star calculations. One should note, however, that our calculations for the bulk viscosity coefficients due to phonon processes are only valid when the phonons are behaving as a fluid, that is, when the mean free path is smaller than the size of a neutron star. In Ref.~\cite{Manuel:2011ed}, we show that this is the case when temperatures are of the order of $T \gtrsim 10^8$ K, the exact temperature depending on the density studied. The bulk coefficients using the  $^1S_0(A)+^3P_2(i)$ neutron gap model are shown on the left panels while those for the $^1S_0(a)+^3P_2(h)$ neutron gap scheme are displayed on the right panels. As previously indicated, the $\zeta_2$ coefficient has the same physical meaning as the bulk viscosity in a normal fluid. We observe that the scaling with temperature of all three coefficients is given by the temperature dependence of the phonon decay rate. As we have seen in Sec.~\ref{subsec:num}, the behavior with temperature of the rate is dominated by the contributions of type I, type II and cross diagrams stemming from the collinear region of the  2$\leftrightarrow$3 processes. In particular, type II terms govern the decay rate and, hence, the bulk viscosity coefficients up to $ T\sim 10^9$K for all the densities studied. On the other hand, each of the three coefficients has a distinct density dependence which results from the different combination of the density--dependent $I_1$ and $I_2$ quantities, according to eqs.~(\ref{bulkstatic}, \ref{coef}). A slightly different behavior with density for the bulk coefficients is also manifest depending on the neutron gap model used. By comparing the results for both neutron gap schemes, we note that all bulk coefficients present bigger values for the  $^1S_0(a)+^3P_2(h)$ case for all densities, partially due to the smaller phonon rates as seen in fig.~\ref{rates}.

\begin{figure}
\begin{tabular}{ccc}
\includegraphics[width=7.5cm,height=6cm]{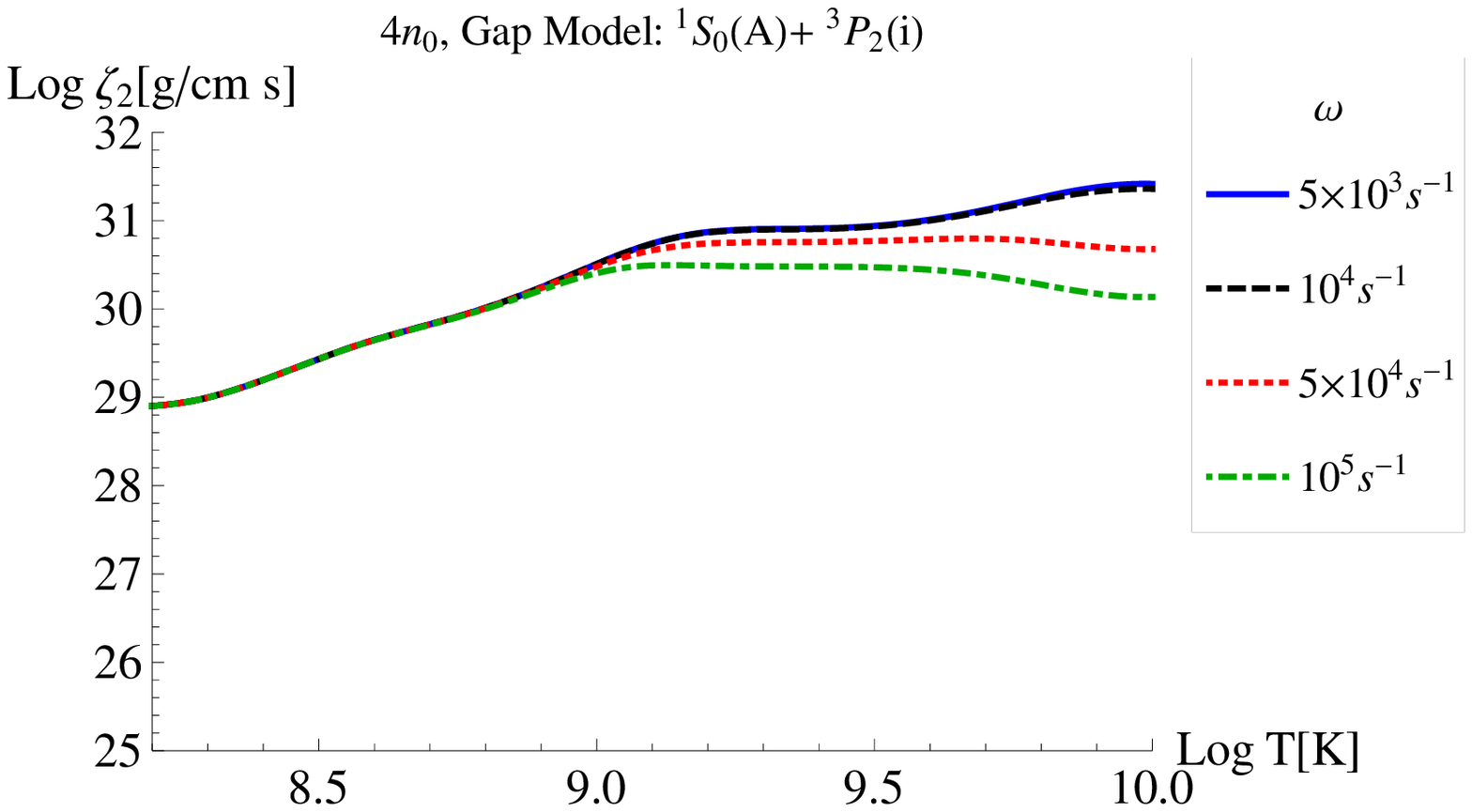} & \includegraphics[width=7.5cm,height=6cm]{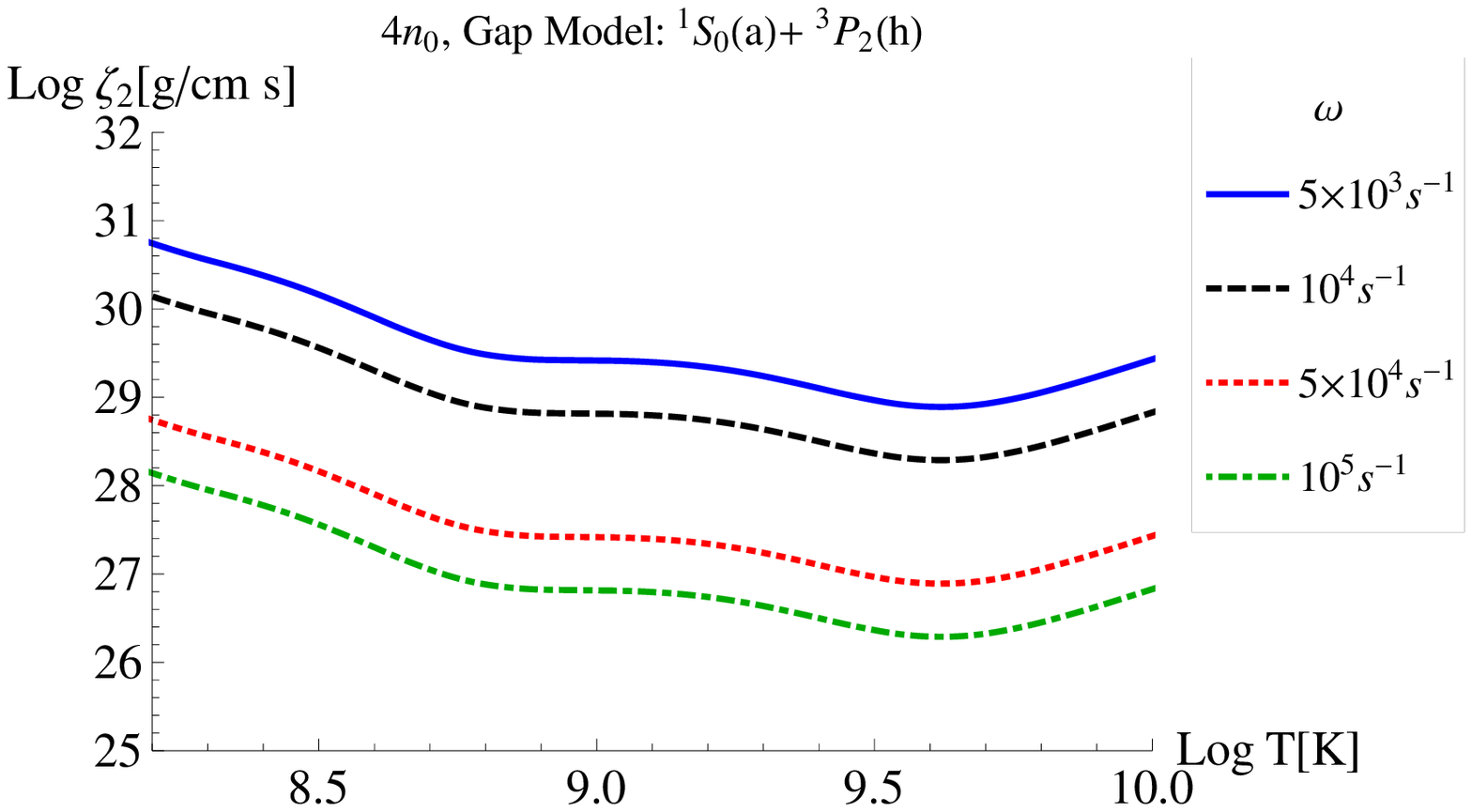}
\end{tabular}
\caption{ $\zeta_2$ frequency--dependent bulk viscosity coefficient as a function of the temperature for $4n_0$ and frequencies between $10^{3}-10^{5}s^{-1}$ for the two neutron gap models.}
\label{zeta2}
\end{figure}

\begin{figure}
\begin{tabular}{ccc}
\includegraphics[width=8cm,height=6cm]{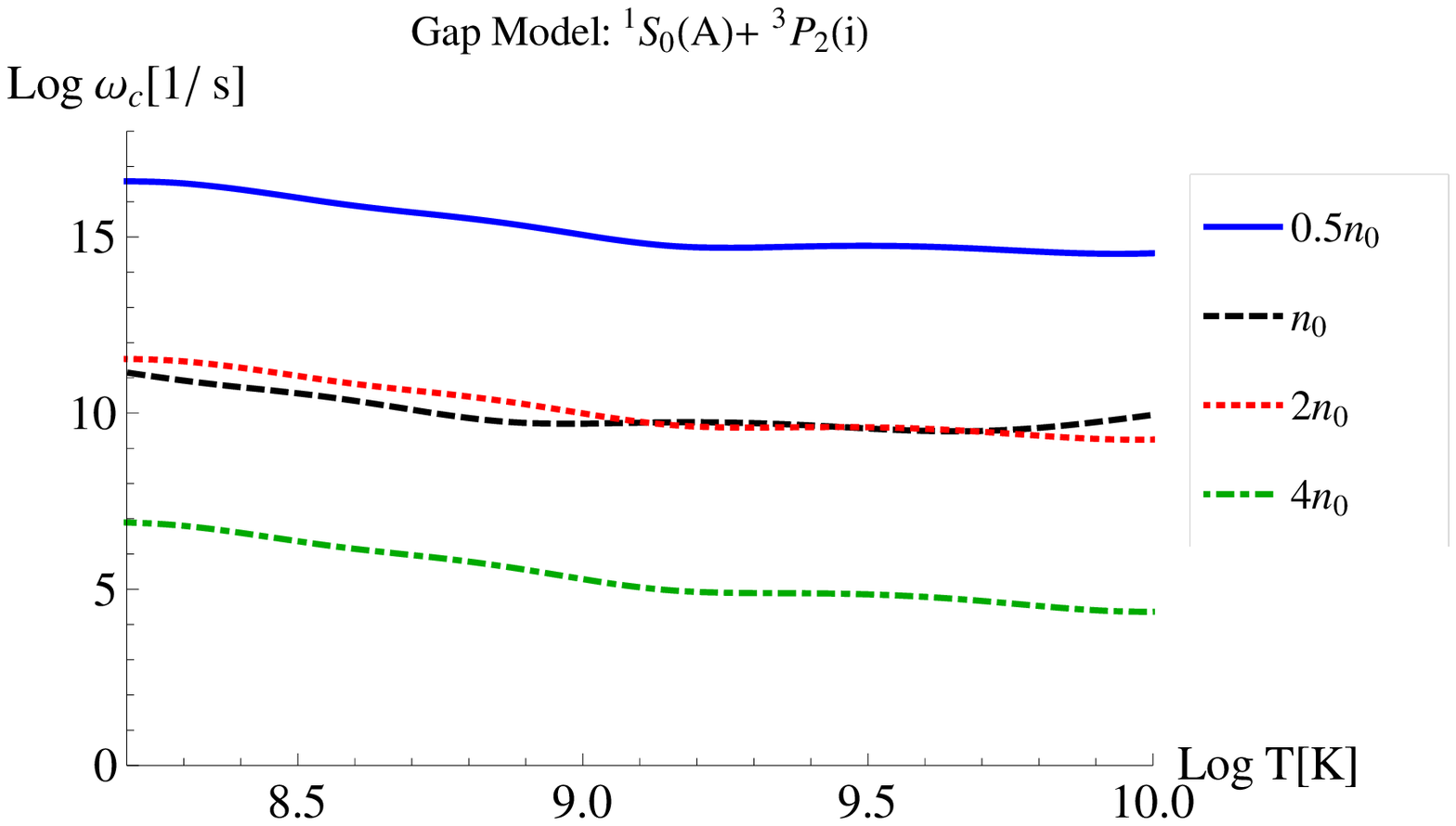}& \includegraphics[width=8cm,height=6cm]{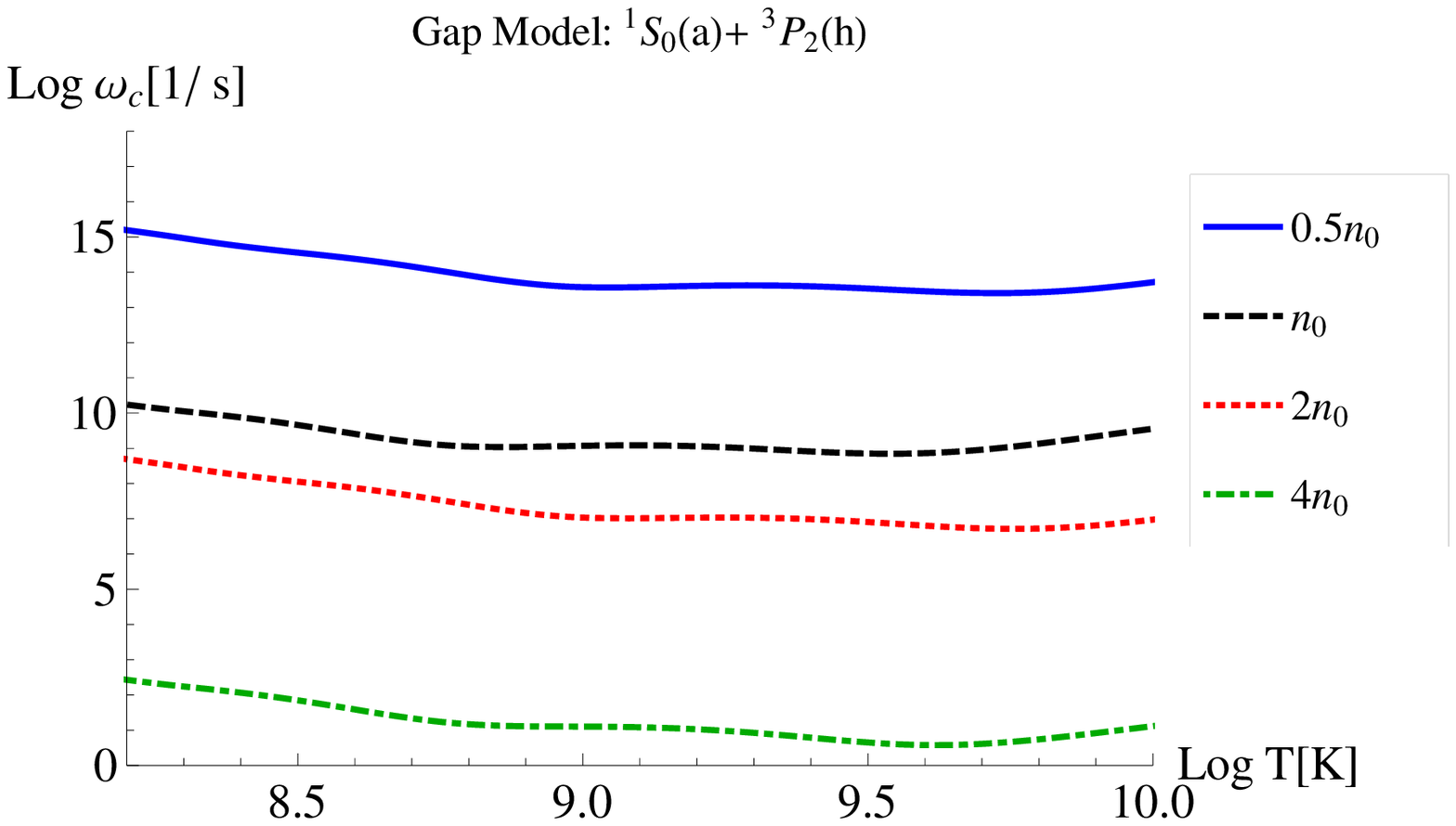}
\end{tabular}
\caption{Characteristic frequency $\omega_c$, defined in eq.~(\ref{char}), at different densities as a function of the temperature for the two neutron gap models.}
\label{frec}
\end{figure}

In fig.~\ref{zeta2} we display the frequency--dependent $\zeta_2$ coefficient  at $4n_0$, which describes the damping of stellar pulsations with typical frequencies of $\omega=10^3-10^5s^{-1}$.  For the two gap models analyzed, we observe some structures at given temperatures for all frequencies under consideration. These are mainly related to the behavior of the $1/(1+(\omega/\omega_c)^2)$ factor with temperature and, in particular, of  the characteristic frequency $\omega_c$. This quantity behaves with temperature as $ \omega_c \sim I_1^2 \ T / \Gamma(T) \sim T^{10}\ T / \Gamma(T)$, where $ \Gamma(T)$ is a complicated function of temperature, as seen in fig.~\ref{pccon} for the relative contributions of the different terms to $ \Gamma(T)$. Thus, the structures seen for each given frequency in fig.~\ref{zeta2} correspond to the variations of $ \Gamma(T)$ with temperature. When using the $^1S_0(A)+^3P_2(i)$ neutron gap model, we observe  on the left-hand side of fig.~\ref{zeta2} that for $\omega= 10^4s^{-1}$  the bulk viscosity is different by more than 10$\%$ from its static value only for $T\gtrsim 10^{10}$K, while for $\omega= 10^5s^{-1}$ the difference is larger than 10$\%$ for $T\gtrsim 10^9$K. In fact, the left panel of fig.~\ref{frec} shows that the static approximation is valid for the studied range of densities and temperatures, with the exception of $4n_0$  and $ T\gtrsim 10^9$K since the characteristic frequency $\omega_c$ becomes comparable to the typical value of the radial pulsations in stars. On the contrary, the $\zeta_2$ coefficient is strongly dependent on the frequency if the  $^1S_0(a)+^3P_2(h)$ model is considered, as seen in the right-hand side of fig.~\ref{zeta2}.  In this case  the characteristic frequency of the right panel of fig.~\ref{frec} becomes similar or even smaller than the typical stellar pulsation frequencies for all temperatures at $n \gtrsim 4n_0$ and, thus, the $\zeta_2$ coefficient is suppressed with increasing frequency, an effect that can be easily inferred from eqs.~(\ref{w-bulks},\ref{char}).  Similar behavior is expected for the frequency--dependent $\zeta_1$ and $\zeta_3$ bulk viscosity coefficients. 

We can now compare our results for the frequency--dependent $\zeta_2$ bulk viscosity coefficient coming from the collisions among superfluid phonons with the contribution stemming from direct Urca \cite{Haensel:2000vz} and modified Urca \cite{Haensel:2001mw} processes for a typical  frequency of $\omega=10^4 s^{-1}$. From the results presented in fig.~2 of Ref.~\cite{Haensel:2000vz} for the contributions of direct Urca processes at $T=10^9$K in non-superfluid matter, we see that the phonon contribution to the bulk coefficient using both neutron gap models is several orders of magnitude larger except in Model II   \cite{Prakash:1988md} around densities of $2n_0$, when the sudden opening of the Urca processes takes place.  In fig.~7 the contributions to the viscosity from direct Urca processes in Model II at density $4n_0$ is plotted as a function of the temperature for non--superfluid and superfluid matter, showing that these contributions are smaller for all temperatures in the range $10^8<T(K)<10^{10}$ although converging in size as the temperature increases. The comparison is similar for modified Urca processes. In fig.~1 of  Ref.~\cite{Haensel:2001mw} these contributions in non--superfluid matter as a function of the density for $T=10^ 9$K are displayed, being much smaller than the phonon contributions considering both neutron gap models except for densities about $2n_0$ in Model II when the Urca processes open up. In fig.~4 the viscosity for Model I  \cite{Prakash:1988md} and $2n_0$ density is plotted as a function of the temperature for both non--superfluid and superfluid matter, the contributions being much smaller at low temperatures and converging to the same order of magnitude at large temperatures. 
Note that the bulk viscosities coming from direct Urca and modified Urca are estimated by taking their values in the normal phase and multiplying them by reduction factors. 
It is well possible that these estimates miss the fact that there are points in the phase space where the neutrons are gapless and might contribute in a much more relevant way to the transport coefficients. Indeed, the gapless neutron modes would only be relevant if there were also gapless proton modes that they could convert to via weak interactions. A much more detailed study of this fact deserves further investigation.

\section{Conclusions}
\label{sec:conc}

We have computed the three bulk viscosity coefficients that appear in the superfluid hydrodynamic equations as arising from the collisions among phonons in superfluid neutron stars. We have presented a detailed analysis of how the phonon dispersion law determines the possible collisional processes relevant for the computation of these transport coefficients, and also how their explicit values depend on the EoS of the nucleonic matter inside the star as well as the neutron pairing gap. However, our method of computation is rather general, and could be used for different superfluid systems, provided they share the same underlying symmetries. Only the knowledge of the EoS of the superfluid and the specific form of the phonon dispersion law would be needed to extract the value of the three bulk viscosity coefficients from our general formulation.  

In this article we have used the APR EoS in a causal form \cite{Heiselberg:1999mq} to describe the $\beta$--stable nuclear matter inside the star, while two very distinct parameterizations of  the sum of the $^1S_0$ and the angle-averaged $^3P_2$ neutron gaps have been considered \cite{Andersson:2004aa}. Whereas the APR EoS is a common benchmark for all the nucleonic EoS, the neutron pairing gaps are still model dependent \cite{Lombardo:2000ec}. Thus, we have employed two very different  neutron gap functions in order to test the model dependence of our results as the exact numerical values for the bulk viscosity coefficients depend unavoidably on this gap. Nevertheless, any future improvement in the determination of the neutron pairing gaps can be easily accommodated in our general scheme. 

Our results indicate that the dominant contribution of type I, type II and cross diagrams to the three bulk viscosity coefficients comes from the collinear regime of the 2$\leftrightarrow$3 processes. In particular, the collinear behavior of the type II terms governs the temperature dependence of the bulk viscosity coefficients up to $T \sim 10^9$K for all the densities studied.  We have also analyzed the frequency--dependent bulk viscosity coefficients as compared to the  static case. We find that it is possible to distinguish between static and frequency--dependent values for densities of $n \gtrsim 4n_0$ depending on the model used for the gap, being the frequency--dependent coefficient suppressed with respect to the static one as the frequency becomes bigger than the characteristic frequency for phonon collisions. Finally we have compared our results with those obtained for the bulk viscosities arising from direct and modified Urca processes  \cite{Haensel:2000vz,Haensel:2001mw}. We conclude that,  at $T\sim 10^9$K and  for typical radial pulsations of the star of  $\omega \sim 10^4 s^{-1}$, phonon collisions give the leading contribution to the bulk viscosities in the core, except for $n \sim 2n_0$ when the sudden opening of the Urca processes take place.  Note that our calculations for the bulk viscosity coefficients due to superfluid phonons  are valid up to $T_c  \sim 1/2 \, \Delta \sim 10^{10}$ K, which is approximately the transition temperature to the normal fluid for the given neutron pairing gaps.
 
Our outcome can be used for studying damping of pulsations in neutron stars and gravitational radiation driven instabilities in rotating neutron stars  \cite{Andersson:2000mf,Lindblom:2001}, or the propagation of the sound waves within the star.  In particular, it could be checked whether the phonon contribution to the bulk viscosities has any impact on the  r--mode instability window of superfluid neutron stars, in the same way as it
has been found that the phonon contribution to the shear viscosity modifies the r--mode instability window \cite{Manuel:2012rd}.

\acknowledgments

This research was supported in part by the Spanish MINECO  under contract FPA2010-16963. LT acknowledges support from the Ramon y Cajal Research Programme from Ministerio de Econom\'{\i}a y Competitividad and from FP7-PEOPLE-2011-CIG under Contract No. PCIG09-GA-2011-291679.

\appendix

\section{Allowed phonon collisions}\label{kin}
\label{app:1}

While using the phonon EFT at LO different phonon collisional processes are kinematically allowed, the physics beyond LO may introduce some kinematical restrictions on some scatterings, as we discuss in this Appendix. Here we will restrict the discussion of the allowed kinematics when a phonon dispersion law of the form given in eq.~(\ref{NLOdisp-law}) is considered. Then one can see that the sign of $\gamma$ plays a crucial role in determining whether some processes are or not possible.

Let us first consider the decay of one to  two phonons, or the reverse process, in general, $1 \leftrightarrow 2$. Labelling the incoming particle as $a$ and the outgoing as $b$ and $c$, energy and momentum conservation imposes 
\bea
E_a & = & E_b+E_c \label{cl12e} , \\
\vec{p}_a & = & \vec{p}_b+\vec{p}_c .
\label{cl12}
\eea
Using the leading order dispersion relation, $E_i=c_s p_i$, in the energy conservation equation (\ref{cl12e}) and then using the momentum conservation relation (\ref{cl12}) to eliminate $p_a$ from the former, we obtain an equality from which we can determine the value for the angle $\theta_{bc}$ between $\vec{p}_b$ and $\vec{p}_c$
\be
p_b p_c\left(1-\cos\left(\theta_{bc}\right)\right)=0\,.
\ee
The solutions  $p_b=0$ or $p_c=0$ are naturally excluded, so one finds $\theta_{bc}=0$. Note that if we eliminate $p_b$ or $p_c$ instead of $p_a$ we will find $\theta_{ac}=\theta_{ab}=0$. We conclude that when all the legs of the 3--phonon vertex are on--shell the associated momenta are collinear. It is possible to compute the correction to $\theta_{bc}$ due to the inclusion of NLO contributions to the dispersion relation. Lets define $\delta\theta$ as a small perturbation on the LO result, $\theta_{bc}=0+\delta\theta_{bc}$. Proceeding as we did before, we use the NLO dispersion relations in the energy conservation relation and eliminate $p_a$ using the momentum conservation. Then expanding both sides of the equation to first order in $\gamma$ and $\delta\theta_{bc}$, we find the NLO correction to $\theta_{bc}$ 
\be
\delta\theta_{bc}=\sqrt{6\gamma}\left(p_b+p_c\right)\,.
\ee
Obviously, for the one to two processes to be kinematically allowed, it is necessary that $\gamma>0$. In other words, when considering the NLO dispersion relations, $\gamma$ must be positive to have all the legs of the 3--phonon vertex on--shell.

A similar analysis can be made for the one to three phonon process, $1 \leftrightarrow 3$. Let us label the incoming particle as $a$ and the outgoing particles as $b$, $c$ and $d$. Solving momentum conservation equation for $p_a$, and introducing it in the energy conservation equation while using the LO dispersion relation, we obtain 
\be
p_b p_c(\cos(\theta_{bc})-1)+p_b p_d(\cos(\theta_{bd})-1)+p_c p_d(\cos(\theta_{cd})-1)=0\,,
\ee
with $\theta_{ij}$ being the angle between the vectors $\vec{p_i}$ and $\vec{p_j}$. Excluding the solutions where two of the momenta are zero, we are left with the following solution
\be
\theta_{bc}=0\,,\quad \theta_{bd}=0\,,\quad \theta_{cd}=0\,.
\label{loang13}
\ee
If instead of solving for $p_a$, we were to solve it for any of the outgoing momenta, this leads to $\theta_{ac}=\theta_{ab}=\theta_{ad}=0$. Now we can study the corrections to the angles due to the NLO corrections. Following the same steps as in the  $1 \leftrightarrow 2$ case, we define a small $\delta\theta$ correction to each one of the angles. We arrive to the following equation
\begin{eqnarray}
&&p_b p_c\delta\theta^2_{bc}+p_b p_d\delta\theta^2_{bd}+p_c p_d\delta\theta^2_{cd} \nonumber \\
&&=6\gamma\left(p_b+p_c+p_d\right)\left(p^2_b p_c+p^2_b p_d+p^2_c p_b+p^2_c p_d+p^2_d p_d+p^2_d p_c+2p_b p_c p_d\right)\,,
\end{eqnarray}
which only has solution for $\gamma>0$ because all terms of the equation are positive. Thus the $1 \leftrightarrow 3$  process is kinematically allowed only when $\gamma>0$. This procedure for analysing the kinematics can be extended to any process of one to $n$ phonons leading to the conclusion that these processes are only allowed for $\gamma>0$.

To find a phonon number changing process that is kinematically allowed for $\gamma<0$ we have to look at the two to three phonon processes, $2 \leftrightarrow 3$. We label the incoming particles $a$ and $b$ and the outgoing $d$, $e$, $f$. If we proceed as previously for the LO dispersion relation we obtain the following relation from the conservation laws
\be
\begin{split}
p_d p_e(1-\cos(\theta_{de}))+&p_d p_f(1-\cos(\theta_{df}))+p_e p_f(1-\cos(\theta_{ef}))=\\
& p_b p_d(1-\cos(\theta_{bd}))+p_b p_e(1-\cos(\theta_{be}))+p_b p_f(1-\cos(\theta_{bf}))\,,
\end{split}
\label{cle23}
\ee
which does not allow to determine any of the variables. Thus the two to three process is kinematically allowed regardless of the sign of $\gamma$.

\section{Phase space integral}
\label{phase-appendix}

We give here some details about the choice of variables in performing the phase space integral of eq.~(\ref {rateIntegral}). We first   integrate over $d^3\vec{p}_b$ and $p_a$ making use of the momentum and the energy Dirac deltas respectively, which reduces the phase space to 
\be
\frac{p^*_{a} p_d p_e p_f dp_f dp_d dp_e d\Omega_a d\Omega_d d\Omega_e d\Omega_f}{2^5(2\pi)^{11}c^6_s\left(p_d(1-\cos(\theta_{ad}))+p_e(1- \cos(\theta_{ae}))+p_f(1-\cos(\theta_{af}))\right)}\,,
\ee
where $d\Omega$ stand for the angular variables, and $p^*_{a}$ is defined as
\be
p^*_{a}=\frac{p_d p_e(1-\cos(\theta_{de}))+p_d p_f(1-\cos(\theta_{df}))+p_e p_f(1-\cos(\theta_{ef}))}{ p_d(1-\cos(\theta_{ad}))+p_e(1-\cos(\theta_{ae}))+p_f(1-\cos(\theta_{af}))}\,.
\ee
We can further simplify the phase space expression by choosing a specific parametrization on the momenta. To define the orientation of a vector we need two angles. We have four vectors, so that amounts to eight angular variables. However, we have the freedom to choose the orientation of the reference frame. This freedom can be used to orientate the $z$ axis along the direction of one of the momenta, and the $zy$ plane to be parallel to the one generated by the same momentum and another of the momenta. We have chosen
\bea
\vec{p}_a&=&p_a\left(0,\,0,\,1\right)\,, \nonumber\\
\vec{p}_d&=&p_d\left(\sin(\theta_d)\cos(\phi_d),\,\sin(\theta_d)\sin(\phi_d),\,\cos(\theta_d)\right)\,, \nonumber \\
\vec{p}_e&=&p_e\left(0,\,\sin(\theta_e),\,\cos(\theta_e)\right)\,, \nonumber \\
\vec{p}_f&=&p_f\left(\sin(\theta_f)\cos(\phi_f),\,\sin(\theta_f)\sin(\phi_f),\,\cos(\theta_f)\right)\,.
\eea
Integrating the ciclic angular variables the phase space is reduced to
\be
\frac{p^*_{a} p_d p_e p_f \sin(\theta_d)\sin(\theta_e)\sin(\theta_f)}{16(2\pi)^{9}c^6_s\left(p_d(1-\cos(\theta_{d}))+p_e(1- \cos(\theta_{e}))+p_f(1-\cos(\theta_{f}))\right)}d\theta_d d\theta_e d\theta_f d\phi_d d\phi_f dp_d dp_e dp_f\,.
\label{phsp}
\ee


%\bibitem{a}
%Author, \emph{Title}, \emph{J. Abbrev.} {\bf vol} (year) pg

%\begin{thebibliography}{99}

%\bibitem{a}
%Author, \emph{Title}, \emph{J. Abbrev.} {\bf vol} (year) pg.

%\bibitem{b}
%Author, \emph{Title},
%arxiv:1234.5678.

%\bibitem{c}
%Author, \emph{Title},
%Publisher (year).

% Please avoid comments such as "For a review'', "For some examples",
% "and references therein" or move them in the text. In general,
% please leave only references in the bibliography and move all
% accessory text in footnotes.

% Also, please have only one work for each \bibitem.

%\end{thebibliography}

\end{document}